\documentclass[12pt]{article}
\usepackage{graphics}
\usepackage{xcolor}
\usepackage{epsfig}
\usepackage{graphicx}
\usepackage{amsmath}
\numberwithin{equation}{subsection}
\usepackage{graphicx}
\begin{document}

\begin{center}
{\Large\bf Elastic Wave Scattering off a Single and Double Array of Periodic Defects} 

\vskip 0.3cm
{\bf O. Haq}${}^{1*}$ and {\bf S.~V.~Shabanov}${}^2$

\vskip 0.3cm
${}^1$ {\it Department of Physics, University 
of Florida, Gainesville, FL 32611, USA}\\
${}^2$ {\it Department of Mathematics, University 
of Florida, Gainesville, FL 32611, USA}\\
${}^*$Corresponding author:\ \ omerhaq1@ufl.edu

\end{center}

\begin{abstract}
Elastic waves scattering off a periodic single and double array of thin cylindrical defects is considered for isotropic materials. An analytical expression for the scattering matrix is obtained by means of the Lippmann-Schwinger formalism
and analyzed in the long wavelength limit using Schloemilch series in order to obtain explicit expressions for the poles of the scattering matrix. 
The latter is then used to prove that for a specific curve in the space of physical and geometric parameters, the scattering is dominated by resonances, and the width of the resonances in the shear mode 
parallel to the cylinders has a global minimum in parameter space. This a feature is not observed in similar photonic or acoustic systems. The resonances in
shear and compression modes that are coupled in the plane perpendicular to the cylinders due to the normal traction boundary condition  are studied for the double array. 
The analytical dependence of the width of these resonances on physical and geometrical parameters
 is exploited to prove the existence of resonances with
the vanishing width, known as Bound States in the Continuum (BSC). 
Spectral characteristics of BSC are explicitly found  in terms of the Bloch phase and 
group velocities of elastic modes.  
 
\end{abstract}

 \section{\Large Introduction}
Elastic and photonic metamaterial structures have been of great interest in the field of wave physics, both mathematically and experimentally, exotic wave properties can be seen across the field of scattering theory from scalar theories such as acoustic wave guides structures \cite{awg}  to photonic structures such as a double arrays of dielectric cylinders \cite{JMP2010},\cite{PRL2008},\cite{SchlomilchEMseries1} that of which has been studied extensively. Elastic metamaterials have also been a field of interest, this is due to the fact that elastic metamaterials structures \cite{MMM1}-\cite{MMM4} have the ability to achieve atypical elastic moduli that can't be achieved in more conventional elastic structures. From a mathematical point of view the structure of the equation differs significantly from the photonic and acoustic counterparts; elastic systems can support both longitudinal and transverse mode as opposed to acoustic and photonic systems which can only support a single polarization. In addition the interface conditions , namely, the normal traction boundary conditions \cite{Landau} require that these polarization couple at the boundary between two different elastic materials. Lastly both polarization propagate at different group velocities. These conditions merit an extensive mathematical analysis on the scattering property of elastic structures with similar geometry's to photonic and acoustic structures which have been studied previously. The focus of this paper will be a single and double array of cylindrical elastic scatters, the photonic counterpart is known for it's ability to support bound states whose frequency lies in the radiation continuum (BSC), these unconventional modes were first discovered by Neumann and Wigner \cite{NW1} in the context of quantum mechanics via an inverse design, the results was further extended and corrected by Stillinger and Herrick \cite{NW2}. 

Since there advent, BSC have been studied in much more depth, there has been a classification structure of different type of BSC, which although may not be mutually exclusive, has revealed the main mechanism that allow for the existence of such modes, this has been detailed extensively in \cite{Hsu}. In addition theses modes have also seen experimental realization in acoustic wake shedding experiments \cite{wakeShedding} and acoustic wave guides \cite{awg}. Layered structures and anistropic acoustic structures have also seen realizations of BSC \cite{anistropic}-\cite{layered}. Of the few types of BSC that have been analyzed, the one of concern in this paper have been commonly refereed to as Fabry-Perrot type BSC, the mechanism behind the formation of these BSC is far field  destructive interference of resonance radiation from two identical resonators through fine tuning of the material, geometrical, and spectral parameters, this results in the localization of radiation in one or more dimensions. In the dielectric double array each array serves as a resonating interfaces and by tuning the distance between the array one can vary the round-trip phase of the propagating modes in order to achieve destructive interference, we will see that there has to be significant modifications to this analysis in order to achieve a BSC in the elastic counterpart. Similar layered periodic structures have had extensive numerical investigations in the context of  fluids (\cite{SchmomilchSeries5}-\cite{SchmomilchSeries15}), acoustics (\cite{Rev7}-\cite{Rev10}), elastics (\cite{Rev1}-\cite{Rev9}), and electromagnetic waves (\cite{JMP2010}-\cite{SchlomilchEMseries1},\cite{subwave}-\cite{SchmomilchSeries6}), atom-photon resonances have even been analyzed in plasmonic-photon cavities \cite{Cited_Me_4} allowing further utilization of quasi-BSC and BSC in quantum technology. Numeric solutions to multipole expansions have been formulated for elastic wave scattering by layered periodic grating structures \cite{Rev1}. Layered structures consisting of both empty and water filled cylindrical inclusions have been discussed in \cite{Rev6} using a similar multipole expansion. These layer by layer methods have even been used to analyze the decoupled out of plane mode for periodic arrays of cylindrical scatterers \cite{Rev7}, \cite{Rev5}, \cite{Rev6}, structures such as these can be used as elastic filters \cite{Rev7},\cite{Rev1},\cite{Cited_Me_1} or waveguides \cite{Rev9}. One can even couple elastic BSC to photonic resonances to exploit  opto-mechanical effect in crystal slabs, such ideas have been investigated using a group theoretic approach \cite{Optomech}. 

Elastic wave phenomena has shown great promise with regards to the rise in popularity of elastic metamaterials. Resonances and BSC have been analyzed in a periodic waveguides connected to side-coupled pillared resonators \cite{Cited_Me_1}, devices such as these can be utilized in order to perform elastic mode conversion. Additionally there has been experimental evidence of BSC in layered structures consisting of solids and liquids \cite{Cited_Me_2}. Topological Bound States have been analyzed in elastic honeycomb plates with pentagonal disinclination \cite{Cited_Me_3}. Analytical and numerical analysis of elastic Fabry Perot BSC in a periodic double array of elastic scatters has already been developed in \cite{Wave_Motion}, this paper is concerned with analyzing the existence of BSC consisting of multiple polarizations in a periodic double array of elastic scatters in the Fabry-Perot limit through means of the partial wave summation for coupled waves. It was shown in this paper that resonances exist in the first open transverse diffraction threshold, further it was shown that one could achieve a BSC consisting of mixed polarizations in the Fabry-Perot limit by tuning the resonance width to zero along certain parameter curves which depend on spectral, geometric, and material parameters. In this paper we are concerned with the existence of BSC in the periodic double array for arbitrary separations, the transcendental equation which determine the resonance frequency and width are derived and evaluated, it is shown that these BSC are preserved for all separations between the two arrays and match the results given in \cite{Wave_Motion} in the Fabry-Perot limit. We analyze the existence of resonances in the asymmetric double array and show that one can tune this resonance to a BSC for certain array offsets. In addition we provide an in depth analysis of the existence of BSC in arbitrary diffraction thresholds, providing the necessary and sufficient conditions for existence of a BSC in higher order diffraction thresholds. Lastly we analyze the ultra-narrow resonances/ quasi-BSC present in the out of plane transverse mode for the single array structure, ineterference between competing elastic effects, namely, variation in density and variation in Lame' coefficients, allows one to tune the resonances width even lower then what is possible in electromagnetic counterpart. This phenomena is not seen in the periodic array of dielectric scatters, it's presence is unique to the elastic wave systems due to the structure of the elastic dipole moment for small elastic scatters. 

%\textcolor{red}{Additionally the elastic single array shows some peculiarity that can only be achieved due to the structure of the elastic dipole moment.\\
%these standing wave modes defy convention wisdom on wave theory and they have been reviewed extensively in a variety of fields (CITE), for the case of the dielectric double array it has been shown that through proper tuning of geometric  parameters, explicitly the width of the double array, that one can tune the far field radiation to zero, hence trapping a mode above the continuum edge, these BSC are refered to as Fabry-Perrot type BSC (CITE). In this paper we will investigate the scattering properties of the elastic single and double array and will show explicitly that one can vary the parameters of the theory in order to tune the far field radiation to zero, hence form a Fabry-Perrot BSC mode composed of mixed polarization as well as provided a complete characterization the scattering properties of the systems for general parameter values. 
%}

The structure of this paper can be broken down into the following make up; the first section is an analysis of the Lippmann Schwinger Integral equation for the elastic single and double array in the limit of long wavelength/ small scatterers, from this analysis we pose the scattering problem and define the diffraction thresholds that are present in periodic scattering structures. In the following sections we define scattering matrices for the single array and formulate the partial wave summation in order to analyze the width of the resonance states in the Fabry-Perot limit and provide the conditions on the material, geometrical and spectral parameters necessary to obtain a BSC. This analytical analysis is complemented by a numerical analysis and fitting in order to validate the existence of BSC in this limit. We finish up this paper by analyzing the existence of BSC in the asymmetric double array for arbitrary separations between the arrays. This analysis is compared to the results of the former section in the proper limit, further confirming our results from the previous section.

%\textcolor{green}{1. Finish the introduction in blue or remove it\\
%2. Explain what is new here in comparison to our paper in Wave Motion, cite it.\\
%3. There were a couple of papers published recently on elastic BSC citing our paper in 
%Wave Motion. Add a comment on these works in regard what is done here. You can find these
%papers via Google Scholar, and I think I sent you a copy at some point.}

\section{Lippmann-Schwinger formalism for a single and double array of elastic cylinders}    
\subsection{Formulation of the problem}

The governing equations for propagation of disturbances in elastic isotropic media
are given by \cite{Landau}:
\begin{equation}\label{2.1.1}
\rho\ddot{u}=\nabla \cdot \sigma(\lambda, \mu, u)
\end{equation}
where $u=u(r,t)$ is the displacement vector field at a point $r$ and time $t$, the double dot
denotes partial derivatives with respect to time, $\nabla$ is the gradient operator
in space, the symmetric 2-tensor $\sigma$ is the stress tensor in the medium. It depends 
on the mass density $\rho=\rho(r)$, Lam\'e coefficients $\lambda=\lambda(r)$ and $\mu=\mu(r)$, and the 
displacement field $u(r,t)$ as
$$
\sigma(\lambda,\mu,u)=\lambda(r)\Big(\nabla \cdot u(r,t))
I+\mu(r)(\nabla u(r,t)+(\nabla u(r,t))^{T}\Big)
$$
where $I$ is the unit tensor. Here and in what follows, the conventional tensor notations are used, e.g.,
$(\nabla u)_{ij}=\nabla_iu_j$ and $(\nabla\cdot \sigma)_i=\nabla_j\sigma_{ji}$, and the superscript $T$
stands for transposition. 
For a homogeneous media, $\rho$, $\lambda$, and $\mu$ are constant, and Eq. (\ref{2.1.1}) admits 
plane wave solutions with three polarization modes. The two transverse modes are known as sheer waves,
and the longitudinal mode is known as compression waves. The transverse and longitudinal waves
propagates with different speeds, denoted $c_t$ and $c_l$, respectively, that depend on the media
parameters. Any inhomogeneity in the media causes scattering of elastic waves. 

An inhomogeneity is described  by the relative mass density and Lam\'e coefficients, denoted 
by $\xi_{\rho,\lambda,\mu}$. For example,
$\xi_\rho(r) =(\rho(r)-\rho_b)/\rho_b$ where $\rho_b$ is the constant mass density of the background
media so that $\xi_\rho(r)\neq 0$ only in regions where the mass density differs from that of the 
background media, and similarly for $\xi_{\lambda,\mu}$. A solution to the scattering problem 
is sought in the form $u(r,t)=e^{-i\omega t}u(r)$ where $\omega$ is the frequency of incident and 
scattered waves. The amplitude $u(r)$ is then shown to satisfy the following equation  
\begin{eqnarray*}
D(\nabla) \cdot u&=&-\omega^{2}\xi_{\rho}u-\nabla \cdot \sigma(\lambda_s,\mu_s, u)\,,\\
D(\nabla)&=& (\omega^{2}+c_{t}^{2}\triangle)I+(c_{l}^{2}-c_{t}^{2})\nabla\nabla\\
\lambda_s&=&(c_{l}^{2}-2c_{t}^{2})\xi_{\lambda}\,,\quad \mu_s=c_{t}^{2}\xi_{\mu}\,.
\end{eqnarray*}
The relative Lam\'e coefficients $\lambda_s$ and $\mu_s$ vanish in the bulk.
They are piecewise continuous, having jump discontinuities
at boundaries of regions occupied by scattering structures.
A solution is sought as a regular distribution that is required to be 
continuous everywhere and having a continuous normal traction $\hat{n}\cdot\sigma$ 
where $\hat{n}$ is a unit normal to a boundary surface of a region occupied a scattering structure (defect).
The latter condition 
results from continuity of the elastic displacement and mechanical equilibrium of the defect respectively, 
and causes a coupling of different elastic polarizations at the interface of a defect. 
A standard approach to solving the problem is based on the Lippmann-Schwinger formalism
in which the differential equation is converted to the integral one using a 
Green's function for the operator $D(\nabla)$ satisfying suitable boundary conditions. 

\begin{figure}[!htb]
\begin{center}
\minipage{0.4\textwidth}
% Double Array
   \includegraphics[width=3in,height=1.5in]{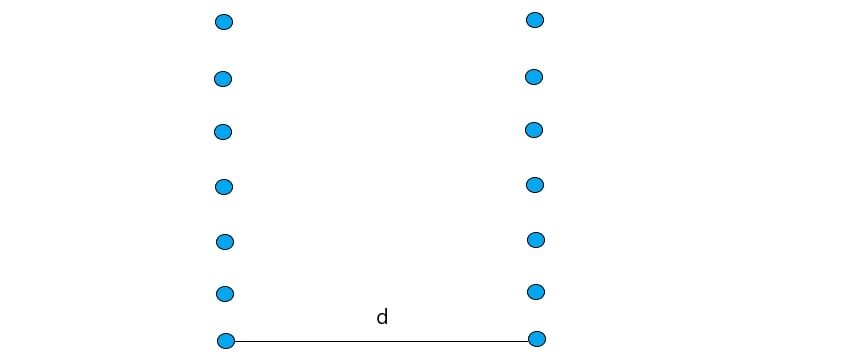}

\endminipage
\end{center}
\caption{\small The double array of cylinders. The cylinders are are infinite and parallel to the ${z}$ axis. 
The ${x}$ axis is horizontal, and the ${y}$ axis is vertical (the array is periodic in this direction). 
The length is measured in units of the period of the array.  The origin is set so that the system 
is symmetric under the reflection $x\to -x$ so that $\Omega(d)=\{r:\mid r-n\hat{y}\pm \frac{d}{2}\hat{x}\mid<R\}$ where $R$ is the radius of the defect and $\hat{x}$ and $\hat{y}$ are unit vectors parallel to the 
coordinate axes.}
\end{figure}
Here the scattering problem is analyzed 
for the system of periodically arranged cylindrical defects (single and double arrays) as depicted in 
Fig. 1. In this case, $u$ is independent of the variable $z$ so that $r=(x,y)$.
One sheer mode is polarized along the $z$ axis, the displacement vector $u$ of the other sheer mode 
and longitudinal mode lie in the $xy$ plane (in-plane modes).
The solution is the sum, 
$$
u(r)=u^{0}(r)+u^{S}(r)\,,
$$
of the scattered wave $u^S$ and
an incident wave that satisfies the associate homogeneous equation
$$
D(\nabla) \cdot u^{0}(r)=0,
$$
which is chosen in the form
$$
u^{0}(r)=-\frac{i}{k_{l}} \nabla(u^{0}_{l}e^{i(k_{l,x}x+k_{y}y)})+
\frac{i}{k_{t}}(\hat{z}\times \nabla ) (u^{0}_{t}e^{i(k_{t,x}x+k_{y}y)})+u^{0}_{t,z}e^{i(k_{t,x}x+k_{y}y)}\hat{z}
$$
where $u_{l}^{0}$, $u_{t}^{0}$ are the (scalar) amplitudes of the in-plane longitudinal and transverse incident 
waves, respectively, while $u^{0}_{t,z}$ is the amplitude of the other sheer mode. 
The magnitude of the wave vector for each polarization satisfies the dispersion relations
$(c_{l}k_{l})^{2}=(c_{t}k_{t})^{2}=\omega^{2}$. So, any plane wave solution is defined 
by its polarization state and a pair of spectral parameters $(\omega^2,k_y)$ (the component $k_x$ is determined
by the dispersion relation). In the asymptotic region $|x|\to \infty$, the scattered wave is also a superposition 
of plane waves with parameters $(\omega^2,k_y')$. Owing to the periodicity of the scattering structure, any solution 
must satisfy the Bloch condition 
$$u(r+\hat{y}) = e^{i k_{y}}u(r)
$$ 
from which it follows that 
$k_y'=k_y+2\pi n$, where $n$ is an integer. The range of $n$ determined by the condition
that $k_{a,x}'$, $a=l,t$, defined by the dispersion relation, is real 

\begin{equation}
k_{a,n,x}^2=\frac{\omega^2}{c_a^2} -k_{y,n}^2>0,
\end{equation}
where 
\begin{equation}
k_{y,n}=k_y+2\pi n. \label{k_yn}
\end{equation}

As a consequence, each incident wave $(\omega^2,k_y)$ can scatter into finitely many {\it open diffraction channels}
defined by the above condition. The number of open channels depends on $\omega^2$.
The radiation continuum for each mode is defined as $\omega^2>c_a^2k_y^2$, and it consists of intervals
in which one or two or three (and so on) diffraction channels are open.

Owing to the Block condition, the scattered field should be sought in the region
$D= \{(x,y): -\infty<x<\infty, -\frac{1}{2}<y<\frac{1}{2}\}$. The scattered wave (in any open diffraction channel)
must carry an energy flux away from the array in the asymptotic region $|x|\to \infty$ and, hence,
satisfy the Sommerfeld radiation condition
\begin{equation}
\nabla_{x}u^{S}_{a,n}(r)\mp ik_{a,n,x}u^{S}_{a,n}(r)\to 0\quad {\rm as}\ x\to \pm\infty
\end{equation}
Given the Bloch condition and the Sommerfeld radiation condition one can deduce the form of the scattered
elastic field in the far field region: 
$$
u^{S}(r) \sim \frac{i}{k_{l}} \sum_{n}c^{\pm}_{l,n} \nabla e^{\pm ik_{l,n,x}+ik_{y,n}y}+
\frac{i}{k_{t}} \sum_{n}\Big[c^{\pm}_{t,n} \hat{z}\times \nabla 
+ c^{\pm}_{z,n} \hat{z} \Big] e^{\pm ik_{t,n,x}+ik_{y,n}y},
$$
the sums are taken over open diffraction channels for each polarization mode $k_{a,n,x}^{2}>0$ labeled by $a=l,t,z$, 
the signs $\pm$ are taken for reflected $(x\to -\infty$) and transmitted $(x\to\infty$) waves.  The scattering 
amplitudes $c_{a,n}^\pm =c_{a,n}^\pm(\omega^2,k_y)$, are functions of the spectral parameters
of the incident wave. The objective is to find these amplitudes.

If the incident wave is set to zero $u^0=0$, then there can still exist solutions $u^S$ that satisfy 
the Bloch and Sommerfeld conditions that are localized, that is, square integrable in the region $D$.
They are called bounded states.
The square integrability implies that bound states cannot have oscillatory behavior in the far
region $x\to \pm \infty$ as propagating waves. Bound states are stationary states, and their energy is not carried away
from the structure in which they are localized. If the frequency $\omega^2$ of a bound state lies below the 
radiation continuum, then this is a regular bound state. If $\omega^2$ lies in an open diffraction 
channel, then such a solution is known as a bound state in the radiation continuum (BSC).
BSCs do not generally exist for any values of physical and geometrical parameters of a scattering structure.
If the scattering amplitudes $c_{a,n}^\pm(\omega^2,k_y)$ exhibit resonances, or poles in the complex    
$\omega^2$ plane with a positive imaginary part (width of the resonance) and with the real part 
being in an open diffraction channel, then the existence of a BSC can be detected by analyzing 
the  width of the resonance as a function of physical and geometrical parameters. If the width 
can be driven to zero by varying these parameters, then there exists a BSC as a resonance 
with the vanishing width. 

Physically, an excited resonance decays in time that is reciprocal of the resonance
width and, hence, the same time is needed to excite the resonance by an incident (incoming) wave. 
The resonance initial energy is carried to 
the asymptotic region by outgoing radiation fields with frequency determined by the real part of the pole. 
If the width can reach zero at certain values of parameters, the resonant 
state can live infinitely long time and becomes a stationary (bound) state. It decouples from the radiation 
continuum. 

In what follows, the resonant 
properties of the scattering amplitudes will be investigated to show that a double array can have 
resonances in open diffraction channels whose width vanishes for certain material, geometrical, and spectral parameters, and the corresponding BSCs will be found. The key 
difference between BSC analytically found in similar electromagnetic systems is that 
the elastic BSC do not have any particular polarization because the longitudinal and transverse polarization 
modes couple at the interface as a result of the normal traction boundary condition. If one makes 
an analogy with Maxwell's theory, then the elastic BSCs found below would correspond to electromagnetic
BSCs in which two transverse polarization modes are coupled and have different dispersions (anisotropic background
media).

\subsection{Lippmann-Schwinger equation}

Suppose that the scatters are homogeneous so that the relative mass density and Lam\'e coefficients
are constant in the region occupied by the scatters so that $\xi_\rho(r)$, $\lambda_s(r)$, and $\mu_s(r)$ are 
piecewise constants and vanish outside of the scatterers.
Then the differential equation for the elastic field satisfying the Sommerfeld condition 
is shown to be equivalent to the the Lippmann-Schwinger integral equation
\begin{equation}\label{LSE}
u_{i}(r)=u_{i}^{0}(r)- \omega^{2} [G_{ij}*(\xi_{\rho}u_{j})](r)-[\nabla_{n}G_{ij}*(\sigma_{jn}(\lambda_s,\mu_s,u))](r)\,.
\end{equation}
where $G_{ij}$ is the Green's function for the operator $D(\nabla)$ that satisfies the Sommerfeld radiation conditions
(its explicit form is given in Appendix A1), and
the star stands for the convolution defined in the distributional sense. 
The support of the region occupied by scatterers is not bounded in the plane
and, hence, the existence of the convolution should be investigated.
It was shown in \cite{JMP2010} that the convolution
$H_{0}^{(1)}(k|r|)*f(r)$,
where $f$ is a regular tempered distribution that satisfies the Bloch periodicity condition
and has support on bounded non-overlapping scatterers
exists in the sense of distributions. Owing to the existence of the convolution, its derivatives 
can be applied to either of the distributions in the convolution in the last term on the right-hand side.

\subsubsection{A solution for a single array in the long wavelength limit}

Here the Lippmann-Schwinger equation is solved in the long wavelength limit, when the wavelength of the 
incident wave is much larger than the radius of cylinders, 
$(\omega R\ll c_{a})$. Consider first the single array case. The centers of cylinder are at $r_n=n\hat{y}$,
where $n$ ranges over all integers.
In this approximation, the scattered wave is a superposition of outgoing 
waves produced point sources located at positions of the scatterers. The strength of point sources
is determined by the Lippmann-Schwinger equation in which the elastic field $u_j$ and the corresponding 
stress tensor $\sigma_{ij}$ can be assumed to have constant values across each scatterer, that is,
$u_j(r)\approx u_j(r_n)$ and $\sigma_{ij}(r)\approx \sigma_{ij}(r_n)$.
Owing to the Bloch condition, $u_{j}(r_n)=e^{ik_yn}u_j(0)$. Therefore, the induced sources are determined
by unknowns $\bar{u_{i}}=u_{i}(0)$ and 
$\bar{\sigma}_{ij}=\sigma_{ij}(\lambda_s,\mu_s,u)(0)$. Their values 
are determined by the Lippmann-Schwinger equation at $r=0$. The technical details as well as the justification 
of this approximation in the case of elastic waves are given in Appendix A2. The solution is given by
\begin{eqnarray}\nonumber
u_{i}(r)&=&u_{i}^{0}(r)-\omega^{2} \xi_{\rho} \bar{u_{j}} \Pi_{ij}(r)-\bar{\sigma}_{jl}\Pi_{ij,l}(r)\,,\\
\label{PIij}
\Pi_{ij}(r) &=& \sum_{n}e^{ik_{y}n} \int_{\mid r' \mid<R}d^{2}r' G_{ij}(r-n\hat{y}-r')\,,\\ \nonumber
\Pi_{ij,n\cdots l}(r)&=&(\nabla_n\cdots \nabla_l)\,\Pi_{ij}(r)\,.
\end{eqnarray}
The right-hand side of the first equation is the convolution in the Lippmann-Schwinger 
equation in the approximation of $u_i(r)$ and $\sigma_{ij}(r)$ by constant values on the scatterers. 
Here and in what follows $\xi_\rho$, $\lambda_s$, and $\lambda_s$ stand
for the constant values of $\xi_\rho(r)$, $\lambda_s(r)$, and $\mu_s(r)$ on the scatterers.
As noted the far field behavior is determined by the field and the change of stress per unit background density at the centers of the defects. The first equation is used to evaluate the corresponding $\sigma_{ij}(r)$. 
Next, it is demanded that the field $u_j(r)$ and the normal traction $(\hat{n}\cdot\sigma)_j$ are continuous 
at $r=0$. These continuity conditions at $r=0$ in the leading order of a small parameter $\epsilon = \pi R^{2}$
gives
a system of equations relating the field and change of stress per unit density at the center of the defect to the incident values of these quantities evaluated at the center of the defect:
\begin{eqnarray}\label{2.2.3}
(1+\omega^{2}\xi_{\rho}\ \Pi_{ii}(0))\bar{u}_{i}+\bar{\sigma}_{jl}\Pi_{ij,l}(0)&=&\bar{u}_{i}^{0} \\
\label{2.2.4}
\bar{\sigma}_{lj}+\bar{\sigma}_{nm}K_{nmlj}^{(1)}(0)+\bar{u}_{m}K_{mlj}^{(2)}(0)&=&\bar{\sigma}^{0}_{lj}
\end{eqnarray}
where $\bar{u}_{i}^{0}$ and $\bar{\sigma}^{0}_{lj}$ are the incident field and its stress tensor
at $r=0$, respectively, no summation over $i$ is implied in $\Pi_{ii}(0)$, and 
\begin{eqnarray*}
K_{nmlj}^{(1)}(r)&=&\lambda_s\Pi_{pm,np}(r)\delta_{lj}
+\mu_s[\Pi_{mj,ln}(r)+\Pi_{ml,jn}(r)]\,,\\
K_{mlj}^{(2)}(r)&=&\omega^{2}\xi_{\rho}\{\lambda_s\Pi_{pm,p}(r)\delta_{lj}
+\mu_s[\Pi_{mj,l}(r)+\Pi_{ml,j}(r)]\}
\end{eqnarray*}
At this point it should be emphasized that unlike the electromagnetic or acoustic case, the induced sources are defined by $\bar{u_{i}}$ and $\bar{\sigma}_{ij}$, which is a new feature unique to elastic theory. Equations (\ref{2.2.3}) and (\ref{2.2.4}) can be cast in the matrix form:
\begin{equation}\label{Kv}
K(\omega^{2}, k_{y})\bar{v}=\bar{v}^{0}\,,
\end{equation}
where the column vectors are defined by
\begin{eqnarray*}
\bar{v}^{0}&=&\langle \bar{u}^{0}_{x}, \bar{\sigma}^{0}_{xy}, \bar{u}^{0}_{y}, \bar{\sigma}^{0}_{xx}, \bar{\sigma}^{0}_{yy}, \bar{u}^{0}_{z}, \bar{\sigma}^{0}_{zy}, \bar{\sigma}^{0}_{zx}\rangle^{T}\,,\\
\bar{v}&=&\langle \bar{u}_{x}, \bar{\sigma}_{xy}, \bar{u}_{y}, \bar{\sigma}_{xx}, \bar{\sigma}_{yy}, \bar{u}_{z}, \bar{\sigma}_{zy},\bar{\sigma}_{zx}\rangle^{T}
\end{eqnarray*}
The $8\times 8$ matrix $K(\omega^{2},k_{y})$ is block diagonal. Its blocks are
\begin{eqnarray*}
 K_{y}(\omega^{2},k_{y}) &=&\begin{pmatrix}
1+\omega^{2}\xi_{\rho}\Pi_{yy}(0) & \nabla_{x}\Pi_{xy}(0) & \nabla_{y}\Pi_{yy}(0) \cr
K_{yyy}^{(2)}(0) & K_{xxyy}^{(1)}(0) & 1+K_{yyyy}^{(1)}(0)  \cr
K_{yxx}^{(2)}(0) & 1+K_{xxxx}^{(1)}(0) & K_{yyxx}^{(1)}(0)\cr
\end{pmatrix}\,,\\
 K_{x}(\omega^{2},k_{y}) &=&\begin{pmatrix}
1+\omega^{2}\xi_{\rho}\Pi_{xx}(0) & \nabla_{x}\Pi_{xy}(0)+\nabla_{y}\Pi_{xx}(0) \cr
K_{xxy}^{(2)}(0) &1+ K_{xyxy}^{(1)}(0)+ K_{yxxy}^{(1)}(0)
\end{pmatrix}\,,\\
 K_{z}(\omega^{2},k_{y}) &=&\begin{pmatrix}
1+\omega^{2}\xi_{\rho}\Pi_{zz}(0) & \nabla_{y}\Pi_{zz}(0) \cr 
K_{zyz}^{(2)}(0) &1+ K_{yzyz}^{(1)}(0) 
\end{pmatrix}
\end{eqnarray*}
and there is a $1\times 1$ block. The equation is decoupled into four matrix equations:
\begin{eqnarray}\label{2.2.5}
K_{y}(\omega^{2},k_{y})\langle \bar{u}_{y},\bar{\sigma}_{xx},\bar{\sigma}_{yy} \rangle^{T}&=&\langle \bar{u}^{0}_{y},\bar{\sigma}^{0}_{xx},\bar{\sigma}^{0}_{yy} \rangle^{T}\,, \\
\label{2.2.6}
K_{x}(\omega^{2},k_{y})\langle \bar{u}_{x},\bar{\sigma}_{xy} \rangle^{T}&=&\langle \bar{u}^{0}_{x},\bar{\sigma}^{0}_{xy} \rangle^{T}\,, \\
\label{2.2.7}
K_{z}(\omega^{2},k_{y})\langle \bar{u}_{z},\bar{\sigma}_{zy} \rangle^{T}&=&\langle \bar{u}^{0}_{x},\bar{\sigma}^{0}_{zy} \rangle^{T}\,. \\
\label{2.2.8}
(1+K_{xzxz}(0))\bar{\sigma}_{xz}&=&\bar{\sigma}^{0}_{xz}\,.
\end{eqnarray}
The S-matrix coefficients are given by linear combination of the components of $\bar{v}$, hence 
its poles are determined by analytic properties of the inverse of
$K(\omega^{2},k_{y})$ in the complex $\omega^{2}$ plane. In particular, positions of
bound states and resonances are roots of the equation $\det[K(\omega^{2},k_{y})]=0$. 

\subsubsection{The double array case}

For a double array, the origin of the coordinate system is set so that the system is symmetric under 
reflection in the $yz$: $x\to -x$, as shown in Figure 1. The centers of scatterers are 
positioned at $r_n^\pm =\pm \frac d2\,\hat{x}+\hat{y}n$ where $n$ ranges over all integers and $d$ is the distance 
between the arrays. In the long wavelength approximation, the convolution in the Lippmann-Schwinger equation
is computed in the same way as in the single array case using the values of $u_j$ and $\sigma_{ij}$ at the 
centers of the cylinders. The Bloch condition shows that these values are determined only by their values 
at $r_0^\pm$, that is,  by
$\bar{u_{i}}^{\pm}=u_{i}(r_0^\pm)$ and $\bar{\sigma}_{ij}^{\pm}=\sigma_{ij}(\lambda_s, \mu_s, u)(r_0^\pm)$.
 
Let $p=\pm$ be the parity index and, for brevity, $r_\pm =r_0^\pm$. Then 
the solution in the long wavelength limit has the form
$$
u_{i}(r)=u_{i}^{0}(r)- \sum_p\Big[\omega^{2} \xi_{\rho}  \bar{u}_{j}^{p} \Pi_{ij}(r-r_p)+
\bar{\sigma}_{jl}^{p}\Pi_{ij,l}(r-r_p)\Big]
$$
Just like in the case of a single array, this expression is used to find the stress tensor $\sigma_{ij}(r)$.
Next the continuity conditions at $r=r_p$ are used to obtain the equations for the unknowns $\bar{u}_j^p$ and 
$\bar{\sigma}_{ij}^p$:
\begin{eqnarray} \label{2.2.10}
(1+\omega^{2}\xi_{\rho}\ \Pi_{ii}(0))\bar{u}^{\pm}_{i}+\omega^{2}\xi_{\rho}\bar{u}^{\mp}_{j}
\Pi_{ji}(r_\pm)+\bar{\sigma}_{jl}^{\pm}\Pi_{ij,l}(0)+\bar{\sigma}_{jl}^{\mp}\Pi_{ij,l}(r_\pm)=
u_i^{0}(r_\pm) \\ \label{2.2.11}
\bar{\sigma}_{ij}^{\pm}+\bar{\sigma}_{nm}^{\pm}K_{nmij}^{(1)}(0)+\bar{\sigma}_{nm}^{\mp}
K_{nmij}^{(1)}(r_\pm)+\bar{u}_{m}^{\pm}K_{mij}^{(2)}(0)+\bar{u}_{m}^{\mp}K_{mij}^{(2)}(r_\pm)=\sigma_{ij}^0(r_\pm)
\end{eqnarray}
where $\sigma_{ij}^0(r_\pm)=\sigma_{ij}(\lambda_s,\mu_s, u^0)(r_\pm)$ is the stress tensor for the incident field
at $r=r_\pm$. Each of the above equations stands for two equations, one for the top parity indices in each term
and another for the bottom parity indices in each term. 

Equations (\ref{2.2.10}) and (\ref{2.2.11}) can be cast in the matrix form
(\ref{Kv}), where the vector $v$ has 16 components that are 16 unknowns $\bar{u}_j^\pm$ and 
$\bar{\sigma}_{ij}^\pm$, $v^0$ is a 16-vector with components being the corresponding components 
of  $u_i^{0}(r_\pm)$ and $\sigma_{ij}^0(r_\pm)$, and the matrix $K(\omega^2,k_y)$ becomes a $16\times16$ matrix. 
 The system has BSCs
if the associated homogeneous equation (when $v^0=0$ (no incident wave)) has a non-trivial solution at some
positive $\omega^2>0$ that
lies above the continuum threshold (in an open diffraction channel). This means
that the matrix $K$ is singular at such $\omega^2$ and
$\det K(\omega^2,k_y)=0$. The latter equation has no such solutions 
for a single array (in the approximation used) but such solutions do exist
for the double array. A simple physical argument to prove this assertion is to note 
that, owing to the normal traction boundary conditions, the shear mode polarized along the cylinders is decoupled from 
the shear mode polarized in the $xy$ plane and the compression mode (the latter two are coupled through 
the boundary conditions). Therefore the Lippmann-Schwinger equation 
is decoupled into two independent equations 
for $u_z(r)$ and $u_i(r)$, $i=x,y$, referred to in what follows as an {\it out-of-plane mode} and {\it in-plane modes},
respectively.
If $\lambda_s=\mu_s=0$, then
the scattering problem for the out-of-pane mode $u_z$ is identical to that for scattering of electromagnetic waves
on a double array of dielectric cylinders where the incident wave is polarized parallel to the cylinders and 
$\xi_\rho$ plays the role of a relative dielectric permittivity. This system is known 
to have BSCs in multiple open diffraction channels \cite{JMP2010}. 

Our next goal is show that the system also has BSCs for the in-plane modes that are coupled 
via the normal traction boundary condition. Such BSCs do not have any specific 
polarization and do not admit any direct analogy with electromagnetic scattering systems 
with transitional symmetry as electromagnetic waves do not have 
longitudinal polarization (in contrast to compression elastic waves). 
Unfortunately, the equation $\det K(\omega^2,k_y)=0$ is not analytically tractable for general parameters of the theory and only numerical methods apply. However, the equation 
can be analyzed in the so-called Fabry-Perrot limit and has been used to find BSC in similar electromagnetic system \cite{PRL2008}.
In this limit, the distance $d$ between the arrays is assumed to be much larger than the wavelength of the incident 
wave $(\omega d \gg c_{a})$. The idea is to show that the scattering on a single array is resonance-dominated
(a background scattering can be neglected). If the distance between the arrays is large enough, then the scattering 
on the double array is dominated by resonances of each array that are coupled only via propagating modes, while 
evanescent fields generated by each array can be neglected in the vicinity of the other array because evanescent 
fields decay exponentially with increasing the distance from the array. As shown in \cite{PRL2008}, there are 
quantized values of the distance at which a single scattering mode of frequency $\omega$ in an open 
diffraction channel can be trapped between the arrays, with the arrays
acting like perfect mirrors, forming a BSC. The distances at which BSC can be formed are determined by 
the condition that the interference of waves multiply scattered from each array is perfectly destructive in the 
asymptotic region $|x|\to \infty$, meaning that the amplitude of outgoing waves vanishes (the space between 
the arrays becomes a perfect resonator). This argument 
cannot be immediately extended to the elastic case because the scattered field has shear and compression 
modes with different dispersion  relations and, hence, the destructive interference condition is generally 
different for each mode, while the modes are coupled at the scatterers and are both present in the scattered wave. 
Nevertheless, it will be shown that such BSCs do exist not only in one open diffraction channel but also 
in multiple open channels, and they also exist if one array is shifted relative to the other in the $y$ direction 
so that the parity symmetry of the system is broken.

\section{ The Fabry-Perot approximation}

Resonance scattering properties of the single and double array system will be analyzed in this section.
In doing so, square integrable solutions to the homogeneous 
 Lippmann-Schwinger equation $(u^{0}(r) = 0 )$ (or Siegert states) will be studied. 
They occur at generally complex-valued solutions $\omega^2$ to the equation $\det{K(\omega^{2},k_{y})}=0$
and correspond to resonances in the scattered field, the real part of $\omega^2$ defines the position 
(squared frequency) of the resonance (in an open diffraction channel), 
and the imaginary part determines the width of the resonance.
The effects of varying material coefficients, polarization mixing at the interfaces, and different dispersions among the polarizations on resonances will be investigated.
In particular, it will be shown that the width of this resonances has a minimum in parameter space, this feature is absent in the electromagnetic counterpart. Finally, BSCs as resonances with the vanishing width are analyzed
for a double array using the Fabry-Perrot approximation in which the distance between the arrays is much
larger than the wavelength. Analytic solutions for BSCs that contain coupled shear and compression modes 
will be obtained. 

\subsection{Spectral range and the S-matrix} 

Throughout this section the analysis is carried out in the case when each of the elastic modes
has only one open diffraction channel.  Recall that the diffraction thresholds for each mode are defined 
by $\omega_{a,n}^{2}=c_a^{2}k_{y,n}^{2}$ with $n$ being an integer. So, the range of spectral parameters
is restricted as
\begin{eqnarray}\label{eq:spectral range}
\omega_{l,0}^{2}<&\omega^{2}&<\omega_{t,-1}^{2}\,\\ \nonumber
0<&k_{y}&<\frac{2\pi \alpha}{1+\alpha}
\end{eqnarray}
where $\alpha = \frac{c_{t}}{c_{l}}<\frac{1}{\sqrt{2}}$, \cite{Landau}, the upper bound on $k_{y}$ is necessary so that the second open transverse channel lies above the first longitudinal channel, $(\omega_{l,0}^{2}<\omega_{t,-1}^{2})$. Due to the parity symmetry, $(x \rightarrow -x)$, $k_{y}$ can be taken strictly positive. 
In what follows it is also assumed that $k_y$ is bounded from below by some positive threshold value 
in order to keep the diffraction thresholds from merging, for $k_{y}$ below this lower bound the problem becomes indistinguishable from normal incidence.  A discussion of the normal incidence is postponed until the end of the paper.

Let us define the S-matrix. It is assumed that an incident plane wave is propagating in the direction 
of increasing $x$,
$$
u^0(r)=-\frac{i}{k_{l}} \nabla(u^{0}_{l}e^{i(k_{l,x}x+k_{y}y)})+
\frac{i}{k_{t}}(\nabla \times \hat{z}) (u^{0}_{t}e^{i(k_{t,x}x+k_{y}y)})+u^{0}_{t,z}e^{i(k_{t,x}x+k_{y}y)}\hat{z}
$$
where $u^0_l$ is the amplitude of the compression mode, $u^0_t$ is the amplitude of the shear mode 
in the $xy$ plane (the latter modes are the in-plane modes), and $u^0_{t,z}$ is the amplitude of the 
out-of-plane shear mode.
For $x \rightarrow+\infty$ the solution to the Lippmann-Schwinger equation must have the form
$$
u(r)\sim -\frac{i}{k_{l}} \nabla(u^{T}_{l}e^{i(k_{l,x}x+k_{y}y)})+\frac{i}{k_{t}}(\nabla \times \hat{z}) (u^{T}_{t}e^{i(k_{t,x}x+k_{y}y)})+u^{T}_{t,z}e^{i(k_{t,x}x+k_{y}y)}\hat{z},
$$
where $u^T_{l,t}$ are the amplitudes for the transmitted in-plane modes, and $u^{T}_{t,z}$ is the amplitude of the transmitted out-of-plane
mode. Similarly,
for $x \rightarrow-\infty$, the solution reads
$$
u(r) \sim u^0(r)
-\frac{i}{k_{l}} \nabla(u^{R}_{l}e^{-i(k_{l,x}x-k_{y}y)})-
\frac{i}{k_{t}}(\nabla \times \hat{z}) (u^{R}_{t}e^{-i(k_{t,x}x-k_{y}y)})
+u^{R}_{t,z}e^{-i(k_{t,x}x-k_{y}y)})\hat{z}.
$$
The transmission and reflection amplitudes are linear combinations 
the incident amplitudes $u^0_{l,t}$ and $u^0_{t,z}$ with the coefficients being the S-matrix elements.
They can be extracted from the far field behavior of the solution to the Lippmann-Schwinger equation.
In particular, for a single array, this is done by
the Bloch wave expansion of $\Pi_{ij}(r)$ as $\mid x \mid \rightarrow +\infty$. 
The result for two open diffraction channels (one transverse and one longitudinal) is given by:
\begin{eqnarray*}
u_{i}(r) &\sim& u_{i}^{0}(r) +\frac{i \epsilon}{2c_{t}^{2}k_{t,x}}
(\omega^{2} \xi_{\rho}\bar{u}_{j}+\bar{\sigma}_{jl}\nabla_{l})\Big(\delta_{ij}+\frac{\nabla_{i} \nabla_{j}}{k_{t}^{2}}\Big)e^{i(k_{t,x} \mid x \mid +k_{y}y)}\\
&&-\frac{i \epsilon}{2c_{l}^{2}k_{l,x}}(\omega^{2} \xi_{\rho}\bar{u}_{j}+\bar{\sigma}_{jl}\nabla_{l})
\Big(\frac{\nabla_{i} \nabla_{j}}{k_{l}^{2}}\Big)e^{i(k_{l,x} \mid x \mid +k_{y}y)},
\end{eqnarray*}
where $\epsilon=\pi R^2$ (the cross section area of the scatterers, $R\ll1$). 
Comparing this expression with the stated asymptotic form of the solution, it is inferred that
for the in-plane modes
\begin{eqnarray}
\label{3.1.2}
u^{T}_{t}&=& u_{t}^{0}+\frac{k_{t,x}}{k_{t}}\, w_{+,y}^{(t)}
-\frac{k_{y}}{k_{t}}\, w_{+,x}^{(t)},\\
\label{3.1.3}
u^{T}_{l}&=& u_{l}^{0}+\frac{k_{y}}{k_l}\,w_{+,y}^{(l)}
+\frac{k_{l,x}}{k_{l}}\,w_{+,x}^{(l)}\,,\\
\label{3.1.4}
u^{R}_{t}&=& \frac{k_{t,x}}{k_t}\,w_{-,y}^{(t)}
+\frac{k_{y}}{k_{t}}\,w_{-,x}^{(t)}\,,\\
\label{3.1.5}
u^{R}_{l}&=& \frac{k_{y}}{k_l}\,w_{-,y}^{(l)}-\frac{k_{l,x}}{k_{l}}\,w_{-,x}^{(l)}\,,
\end{eqnarray}
and for the out-of-plane mode
\begin{align}
\label{3.1.6}
u^{T}_{t,z}&=u^{0}_{t,z}+w_{+,z}^{(t)},\\
\label{3.1.7}
u^{R}_{t,z}&=w_{-,z}^{(t)}.
\end{align}
where 
$$
w_{\pm,j}^{(a)}=\frac{i\epsilon}{2c_{a}^{2}}\Big(\frac{\omega^{2}\xi_{\rho}}{k_{a,x}}\, \bar{u}_{j}
+\frac{ik_{y}}{k_{a,x}}\,\bar{\sigma}_{yj}\pm i\bar{\sigma}_{xj}\Big)
$$
Note that the S-matrix is block-diagonal. The in-plane and out-of-plane modes are decoupled.

The energy flux carried by a solution to the elastic equations (\ref{2.1.1}) is defined by:
$$
J_{i}(r,t)=-{\rm Re}\,\{ \dot{u}^{*}_{j}(r,t)\sigma_{ji}(\lambda_{0},\mu_{0},u)(r,t) \}.
$$
where the star $*$ stands for complex conjugation. In particular, the energy flux across a unit area
in the direction of propagation carried by a plane wave of a polarization mode is proportional
to $c_a|u_a|^2$ where $c_a$ is the group velocity and $u_a$ is the amplitude of the wave.  
Using the asymptotic form of the scattered field, the energy transmission and reflection coefficients
are obtained:
\begin{eqnarray*}
t_{E}(\omega^{2},k_{y})&=&\frac{c_{t}\mid u^{T}_{t} \mid^{2}+c_{l}\mid u^{T}_{l} \mid^{2}+c_{t}\mid u^{T}_{t,z} \mid^{2}}{c_{t}\mid u^{0}_{t} \mid^{2}+c_{l}\mid u^{0}_{l} \mid^{2}+c_{t}\mid u^{0}_{t,z} \mid^{2}}\,,\\
r_{E}(\omega^{2},k_{y})&=&\frac{c_{t}\mid u^{R}_{t} \mid^{2}+c_{l}\mid u^{R}_{l} \mid^{2}+c_{t}\mid u^{R}_{t,z} \mid^{2}}{c_{t}\mid u^{0}_{t} \mid^{2}+c_{l}\mid u^{0}_{l} \mid^{2}+c_{t}\mid u^{0}_{t,z} \mid^{2}}
\end{eqnarray*}
So, any pole in the complex $\omega^2$ plane in the S-matrix elements appears as a resonance 
in the reflection and transmission coefficients as functions of the incident wave frequency.

\subsection{Single Polarization BSC}

As noted, the scattering matrix for the out-of-plane mode is identical the scattering matrix for an analogous 
electromagnetic system studied in \cite{PRL2008,JMP2010} if $\xi_{\mu}=\xi_{\lambda}=0$. Here the effects
of non-zero relative Lam\'e coefficients are investigated. In particular, the existence of BSC in a double array 
will be reexamined. Using the results from the previous sections and the appendix the reflection coefficients are found
for a single array:
\begin{eqnarray*}
r_{t,z}(\omega)&=&\frac{u_{t,z}^{R}}{u_{t,z}^{0}}= \frac{ik_{t,x}\xi_{\mu}}{2}
\frac{\epsilon}{1+c_{t}^{2}\xi_{\mu}\Pi_{zz,xx}(0)}
+\frac{\epsilon\Lambda_z(\omega^2,k_y)}{\det{K_{z}(\omega^{2},k_{y})}}
\\
\Lambda_z(\omega^{2},k_{y})&=&\frac{\omega^{2}\xi_{\rho}\xi_{\mu}}{2k_{t,x}}
\Big(i\Pi_{zz,yy}(0)-ik_{y}^{2}\Pi_{zz}(0)+2k_{y}\Pi_{zz,y}(0)\Big)+
 \frac{i}{2c_{t}^{2}k_{t,x}}(\omega^{2}\xi_{\rho}-c_{t}^{2}k_{y}^{2}\xi_{\mu})\\
\det{K_{z}(\omega^{2},k_{y})}&=&\Big(1+\omega^{2}\xi_{\rho}\Pi_{zz}(0)\Big)
\Big(1+c_{t}^{2}\xi_{\mu}\Pi_{zz,yy}(0)\Big)-c_{t}^{2}\omega^{2}\xi_{\mu}\xi_{\rho}\Pi^2_{zz,y}(0)
\end{eqnarray*}
Let us prove that
the single array has a resonance near the diffraction threshold and extract the standard Breit-Wigner form
of the reflection coefficient form near the resonance. To this end, consider the following parameter curve:
\begin{align}\label{tau}
\frac{\epsilon}{\Delta}=\tau\,,\\
\label{Delta_-1}
\Delta^{2}=\omega_{t,-1}^{2}-\omega^{2},
\end{align}
where $\tau$ is a fixed complex number. Since the real part of $\omega^2$ is close to the diffraction 
threshold, $\tau\sim O(1)$ in the small parameter $\epsilon$.
 Throughout the rest of the paper, this parameter curve will be commonly exploited in order to locate BSC and resonances. The focus will be on the oblique incident case. For small scatters this sets a lower bound on $k_{y}$, as 
noted in the previous subsection,
\begin{eqnarray}\label{eq:kybound}
\frac{\Delta^{2}}{8\pi c_{t}^{2}}<k_{y},
\end{eqnarray}
because
below this bound $\omega_{t,1}^{2} -\omega_{t,-1}^{2}\sim O(\epsilon^{2})$ and the solution becomes indistinguishable from the normal incident case, $k_{y}=0$, to leading order in $\epsilon$. In other words, the diffraction thresholds "merge" and the following analysis will require modification. The case of normal incidence will be discussed at the end of the last section of this paper. Technical details of calculation of $\Pi_{ij}(0)$ and the derivatives
$\Pi_{ij,kl\cdots}(0)$ are given in Appendix A3. Using them and
expanding to the leading order it is inferred that 
\begin{eqnarray*}
\Pi_{zz}(0)&=&-\frac{\tau}{2c_{t}}+\epsilon\beta_{0}(\Delta^2)+\frac{\epsilon\ln(\epsilon)}{4\pi c_{t}^{2}}\,,\\
\Pi_{zz,y}(0)&=&-\frac{i\omega_{t,-1}}{2c_{t}^{2}}\tau+\epsilon\beta_{1}(\Delta^2)\,,\\
\Pi_{zz,yy}(0)&=&\frac{\omega_{t,-1}^{2}}{2c_{t}^{3}}\tau+\epsilon\beta_{2}(\Delta^2)-
\frac{\omega^{2}\epsilon\ln(\epsilon)}{8\pi c_{t}^{4}}\,,
\end{eqnarray*}
where the functions $\beta_{0,1,2}$ are analytic. So, 
the pole is independent of them, and for this reason, the explicit form of the $\beta$'s is omitted
(if so desired, it can be deduced from $\Pi_{ij,kl\cdots}$ given in Appendix A3). 
Next, the determinant  $\det{[K_{z}(\omega^{2},k_{y})]}$ and the reflection coefficients are 
also expanded to the leading order by means of the above equations. 
After some algebraic transformations, the reflection coefficient is reduced to the Breit-Wigner form
\begin{eqnarray*}
r_{t,z}(\omega) &=& -\frac{i\Gamma}{\omega^{2}-\omega_{0}^{2}+i\Gamma}+O(\epsilon)\,,\\
\omega_{0}^{2} &=& \omega_{t,-1}^{2}\Big[1-\frac{\epsilon^{2}\omega_{t,-1}^{2}}{4c_{t}^{2}}
(\xi_{\rho}-\xi_{\mu})^{2}\Big]
+O(\epsilon^{3}\ln{\epsilon})\,,\\
\Gamma &=& \frac{\epsilon^{3}\omega_{t,-1}^{4}(\xi_{\rho}-\xi_{\mu})}{4c_{t}^{2}p_{t,x}}
(k_{y,-1}\xi_{\rho}-k_{y}\xi_{\mu})^{2}+O(\epsilon^{4}\ln{\epsilon})\,,
\end{eqnarray*}
where $k_{y,-1}=k_y-2\pi<0$, and  
\begin{equation}
p_{t,x}=\sqrt{k_{y,-1}^{2}-k_{y}^{2}} \label{p_tx}
\end{equation}
is the $x$-component of the wave vector for the open transverse channel. 
The pole describes a scattering resonance if $\Gamma>0$ so it is necessary that 
$\xi_{\rho}\geq \xi_{\mu}$. As noted earlier, if the resonance width can be driven to zero, then
the corresponding pole can correspond to a BCS. This is not possible for the analogous electromagnetic 
problem \cite{JMP2010}, but it is possible, at least in the leading order, for the elastic case.
First, $\Gamma =O(\epsilon^4\ln\epsilon)$ if  $\xi_{\rho}=\xi_{\mu}$. 
However, the corresponding solution to the Lippmann-Schwinger equation is not square integrable and, hence, unphysical
(it has infinite energy).
Indeed, in the limit $\xi_{\rho}-\xi_{\mu} \rightarrow 0$ and $\Delta \rightarrow 0^{+}$, the asymptotic form of the 
solution is given by
$$
u_{z}(r) \sim \frac{\epsilon \xi_{\rho} \omega_{t,-1}^{2}}{c_{t}} \cdot \frac{e^{-\frac{\Delta \mid x\mid}{c_{t}}}e^{i k_{y,-1} y}}{\Delta}+O(\epsilon)
$$
Its norm is infinite in this limit and, hence, this solution cannot correspond to a physical state.
Second, $\Gamma =O(\epsilon^4\ln\epsilon)$ occurs
 along the parameter curve
$$
\xi_{\rho}=\frac{k_{y}}{k_{y,-1}}\,\xi_{\mu}\,.
$$
In this case, 
$$
\omega_{0}^{2}= \omega_{t,-1}^{2}[1-(\epsilon \pi \xi_{\mu})^{2}]+O(\epsilon^3\ln\epsilon)
$$
and the solution has the form
$$
u_{z}(r)=-\frac{k_{y}}{2\pi}{\rm sign}(\xi_{\mu})e^{-(\frac{\epsilon \pi \omega_{t,-1} \mid\xi_{\mu}\mid}{c_{t}})\mid x \mid}e^{i k_{y,-1} y}\Big(1+\frac{1}{1+\frac{2\pi c_{t}}{\omega_{t,-1}}\,{\rm sign}(\xi_{\mu})}\Big)+O(\epsilon).
$$
where ${\rm sign}(x)=x/|x|$, $x\neq 0$, denotes the sign function.
It is square integrable (normalizable) and, hence, is a physical solution. Unfortunately, it is difficult 
to prove whether there exists a curve in the space of parameters along which $\Gamma=0$ in all orders
of the perturbation theory, which would imply that a single array supports BSC that occur due to a 
fine tuning of the mass density and Lam\'e coefficients.  
It should be emphasized that this state exists only if $\xi_\mu <0$ because in the leading order
$$
\det{K_{z}(\omega^{2},k_{y})}= 1+{\rm sign}(\xi_{\mu})+O(\epsilon)\,.
$$
This implies that the relative mass density must be positive, $\xi_{\rho}>0$,
as follows from the above parameter curve and that $k_{y,-1}<0$.

The observed state gives some insight into the competing effects from variations of density and Lam\'e coefficient. 
In the Lippmann-Schwinger equation $u_i=u_i^0+G_{ij}*P_j$, the vector field $P_j$ 
can be interpreted as the source for the outgoing wave induced by an incident wave. Its $z$ component  is given by
$$
P_{z}(r)=-\rho_{0}\Big(\omega^{2}\xi_{\rho}(r)u_{z}(r)+c_{t}^{2}\xi_{\mu}(r)\Delta u_{z}(r)\Big)
$$
This shows that
the vanishing width in the leading order along the stated curve in the parameter space 
can be explained as a perfect destructive interference (in the leading order of $\epsilon$) 
of outgoing waves produced by
two terms in the source density, one of which is proportional the relative density $(\xi_{\rho})$ and the other to the relative 
shear Lam\'e coefficient $(\xi_{\mu})$. 
This feature is unique to the elastic scattering on a single array 
of cylinders, and it does not exists for similar electromagnetic systems studied earlier \cite{PRL2008}.
From a practical perspective, this artifact can be used to design extremely narrow resonances that 
cannot be achieved in the dielectric single array.

Having shown that the single array has resonances and the scattering is resonance dominated in the leading 
order of $\epsilon$, it is now not difficult to prove that the double array supports BSC at least 
in the Fabry-Perot limit, $\omega d \gg c_{t}$. As already noted, in this limit the reflection coefficient can be computed by 
the partial wave summation as in the Fabry-Perrot interferometer (neglecting the effects of evanescent fields
of each array):
$$
r_{t,z}^{FB}(\omega)=\frac{(1-e^{2ik_{t,x}d})r_{t,z}(\omega)}{1-r_{t,z}^{2}(\omega)e^{2ik_{t,x}d}}.
$$
Using the explicit form of $r_{t,z}$ in the vicinity of a resonance pole, the poles $\omega^2=
\omega_\pm^2-i\Gamma_\pm$ 
of $r_{t,z}^{FB}$ are found:
\begin{eqnarray*}
\omega_{\pm}^{2}&=&\omega_{0}^{2}\mp\Gamma \sin{(k_{t,x}d)}\,,\\
\Gamma_{\pm}&=&[1\pm \cos{(k_{t,x}d)}]\Gamma\,.
\end{eqnarray*}
The parity index $+/-$ corresponds to even/odd parity states.
If the distance between the arrays is tuned 
so that $k_{t,x}d=\pi n$ with $n>0$ being an integer (a large integer 
in the approximation used), then the width of one of the resonances 
vanishes and the width of the other doubles as compared to that for the single array, and in this 
case, the position of the resonances is $\omega^2_\pm =\omega_0^2$. By construction, $\omega_0^2$ lies 
in the open diffraction threshold for the out-of-plane mode and, hence, the resonance 
with the vanishing width is a BSC.

\subsection{Mixed Polarization BSC for arbitrary Lam\'e coefficients}

It was shown that the in-plane scattering modes are coupled through the normal traction boundary 
condition. The corresponding scattering matrix was calculated. Owing to the coupling 
of the polarization modes, the reflection and transmission coefficients form the reflection and 
transmission matrices for a single array. For example, the reflection matrix is defined by
$$
\langle u_{l}^{R},u_{t}^{R} \rangle^{T} = R(\omega^{2},k_{y})\langle u_{l}^{0},u_{t}^{0} \rangle^{T}. 
$$
where the components in the right-hand side are given in (\ref{3.1.4}) and (\ref{3.1.5}). 
The transmission matrix is defined similarly.
Suppose there are two identical arrays
as shown in Figure 1 at a distance $d$ from one another. If the distance is large enough, then 
the reflection matrix of the double array can be computed by summation of all reflected waves produced
by multiple bouncing between the arrays. Since the polarization modes are coupled, and each mode
has its own dispersion relation, the conventional Fabry-Perot summation for a scalar wave 
needs a modification.

Suppose that each interface of a Fabry-Perot interferometer can mix $N$ independent modes of 
an incident wave of frequency $\omega$ that are 
labeled by index $i=1,2,...,N$. Each mode has a group velocity $c_i$. Suppose that
the distance $d$ between the interfaces is much larger than $2\pi \omega/c$ where $c=\max_i\{c_i\}$. 
In this case, the evanescent fields produced by each interface 
can be neglected in the vicinity of the other interface, and reflection and transmission fields for the combined 
structure is determined only by multiple scattering of propagating waves on each interface.
Let the $x$ axis be normal to the interfaces, and $k_{i,x}$ be the $x$ component
of the wave vector of the $i$th mode.  If $R(\omega)$ and $T(\omega)$ are $N\times N$ reflection and transmission 
matrices of the interface, then  the reflection and transmission matrices of the combined 
structure are obtained using the partial wave summation,
\begin{eqnarray*}
	R_{FP}(\omega)&=&R(\omega) +T(\omega)D(\omega,d)R(\omega)D(\omega,d)
[I-(R(\omega)D(\omega,d))^{2}]^{-1}T(\omega),\\
	T_{FP}(\omega)&=& T(\omega)D(\omega,d)[I-(R(\omega)D(\omega,d))^{2}]^{-1}T(\omega).
\end{eqnarray*}
where $I$ is the identity matrix and $D$ is a diagonal matrix,
$$
D(\omega, d)={\rm diag}(e^{ik_{1,x}d},\ e^{ik_{2,x}d},\cdots,\ e^{ik_{N,x}d})
$$
The pole structure of the reflection and transmission matrices of the combined structure 
are defined by zeros of the determinant
\begin{equation}\label{det0}
\det[I-(R(\omega)D(\omega,d))^{2}]=0
\end{equation}
in the complex frequency plane.

Next, suppose that the interface has a resonance. This means that the reflection matrix has a pole
at $\omega^2=\omega_0^2 -i\Gamma$, $\Gamma>0$. Near the pole, the reflection matrix can be written in the form
\begin{equation}
\label{eq:R}
R(\omega) = \frac{\tilde{R}}{\omega^{2}-\omega_{0}^{2}+i\Gamma}+K_{0}(\omega).\
\end{equation}
where $\tilde{R}$ is the residue matrix and the matrix $K_0$ is analytic and describes a background scattering.
If the background scattering can be neglected and 
the resonance is sufficiently narrow ($\Gamma$ is small enough) so that $D(\omega,d)$ can be approximated 
by its value at $\omega_0$, then Eq. (\ref{det0}) is simplified to
$$
\det[(\omega^{2}-\omega_{0}^{2}+i\Gamma)^{2}I-(\tilde{R}D(\omega_{0},d))^{2}]=0.
$$
The left-hand side is a polynomial of degree $N$ in the variable 
$\alpha=(\omega^{2}-\omega_{0}^{2}+i\Gamma)^{2}$. Therefore it has $N$ complex roots, $\alpha=\alpha_j$,
$j=1,2,...,N$, which define positions of new poles, $\omega^2=\omega_0^2-i\Gamma\pm \alpha_j^{1/2}$. 
So, the combined system has $2N$ resonances in general. Some of them can become BSC if their widths
$\Gamma_j^\pm =\Gamma\pm {\rm Im}\,\alpha_j^{1/2}$ can be driven to zero by tuning parameters of 
the system, e.g., the distance $d$.

In particular,
for the in-plane modes in the elastic double array, $N=2$, and the corresponding quadratic equation 
is easy to solve
\begin{equation}
\label{eq:res}
(\omega_{\pm}^{2}-\omega_{0}^{2}+i\Gamma)^{2}=\frac{1}{2}\Big\{{\rm Tr}(\tilde{R}D(\omega_{0},d))^{2}
\pm\sqrt{[{\rm Tr}(\tilde{R}D(\omega_{0},d))^{2}]^{2}-4\det(\tilde{R}D(\omega_{0},d))^{2}}\Big\}
\end{equation}
In what follows, it will be shown that all the assumptions made in the process of deriving
(\ref{eq:res}) are justified for the in-plane modes.

It is assumed that $\xi_{\rho}>0$, otherwise no bound states of any kind can be formed. To find 
the reflection matrix,
equations (\ref{2.2.5}) and (\ref{2.2.6}) must be solved for $\bar{u}_j$ and the stress $\bar{\sigma}_{jk}$. 
The solutions are substituted into (\ref{3.1.4}) and (\ref{3.1.5}) and the matrix elements 
of the reflection matrix are extracted. All calculation should be carried out in the leading order in $\epsilon$.
Technically, the process is similar to the derivation 
of the reflection coefficient for the out-of-plane mode.
 The result reads
\begin{eqnarray*}
R(\omega^{2},k_{y}) &=& \epsilon N(\omega^{2},k_{y})K_{x}^{-1}(\omega^{2},k_{y})Q(\omega^{2},k_{y})+O(\epsilon^2)\,,\\
N(\omega^{2},k_{y}) &=&
\begin{pmatrix}
	-\frac{i\xi_\rho k_{l}}{2} & \frac{k_{y}}{c_{l}^{2} k_{l}} \\ 
	\frac{i \xi_{\rho} k_{y} k_{t}}{2k_{t,x}} &\frac{k_{t,x}^{2}-k_{y}^{2}}{2c_{t}^{2}k_{t}k_{t,x}} 
\end{pmatrix}\,,\\
K_{x}^{-1}(\omega^{2},k_{y}) &=&
\frac{1}{q(\omega^{2},k_{y})}%\[
\begin{pmatrix}
	1+ K_{xyxy}^{(1)}(0)+ K_{yxxy}^{(1)}(0) & -\Pi_{xy,x}(0)-\Pi_{xx,y}(0) \\ 
	-K_{xxy}^{(2)}(0) &1+\omega^{2}\xi_{\rho}\Pi_{xx}(0)  
	\end{pmatrix}\,,\\
Q(\omega^{2},k_{y}) &=&
\begin{pmatrix}
	\frac{k_{l,x}}{k_{l}} & -\frac{k_{y}}{k_{t}} \\ 
	\frac{2ic_{t}^{2}\xi_{\mu}k_{l,x}k_{y}}{k_{l}} & \frac{ic_{t}^{2}\xi_{\mu}(k_{t,x}^{2}-k_{y}^{2})}{k_{t}}  
	\end{pmatrix}\,,\\
q(\omega^{2},k_{y}) &=& \det{K_{x}(\omega^{2},k_{y})}\,.
\end{eqnarray*}
It should be noted that in the frequency range stated earlier, 
the induced source for the scattered wave is proportional to $\bar{u}_{x}$ and $\bar{\sigma}_{xy}$ to 
the leading order, while $\bar{u}_{y},\bar{\sigma}_{xx},$ and $\bar{\sigma}_{yy}$ only contribute to the 
background scattering since
$$
\det{K_{y}(\omega^{2},k_{y})} \sim 1+O(\epsilon)\,.
$$
which implies that their contribution to the induced source is of order $\epsilon$ and, hence, can be neglected near the resonance frequency. 

Next it must be shown that the reflection matrix has a resonance pole whose real part lies in the specified spectral range. As in the previous case, the parameter curve given by (\ref{tau}) and (\ref{Delta_-1}) is used to study
analytic properties of the reflection matrix for small $\epsilon$. 
Using the results on Schl\"omilch series from the Appendix A3 as well as the results from Section 3.2, 
the reflection matrix is proved to have a pole   
at $\omega^2=\omega_0^2-i\Gamma$ where in the leading order in $\epsilon$
\begin{eqnarray*}
\omega_{0}^{2}&=&\omega_{t,-1}^{2}\Big(1-\textstyle{\frac 14} \epsilon^{2}
(\xi_{\rho}-\xi_{\mu})^{2}k_{y,-1}^{2}\Big)\,,\\
\Gamma&=&\epsilon^{3}\tilde{\Gamma}\Big(w_{l}^{2}+w_{t}^{2}\Big)\,,\\
w_{l} &=& \sqrt{p_{l,x}p_{t,x}}\, \Big[2k_{y}\xi_{\mu}
	-k_{y,-1}\xi_{\rho}\Big]\,,\\
	w_{t} &=& (k_{y}^{2}-p_{t,x}^{2})\xi_{\mu}-k_{y}k_{y,-1}\xi_{\rho}\,,\\
	\tilde{\Gamma} &=& \frac{c_{t}^{2}k_{y,-1}^{2}(\xi_{\rho}-\xi_{\mu})}{4p_{t,x}}\,,\\
p_{l,x} &=& \sqrt{\alpha^{2}k_{y,-1}^{2}-k_{y}^{2}}\,,
\end{eqnarray*}

where $p_{t,x}$ is defined in (\ref{p_tx}), and the residue matrix in the leading order is
$$
\tilde{R} = i\epsilon^{3}\tilde{\Gamma}	
\begin{pmatrix}	w_{l}^{2} & -\alpha\sqrt{\frac{p_{t,x}}{p_{l,x}}}\,w_{l}w_{t} \\ 
-\frac{1}{\alpha}\sqrt{\frac{p_{l,x}}{p_{t,x}}}\,w_{l}w_{t} &w_{t}^{2} 
\end{pmatrix}\,.
$$
As in the last section the case $\xi_{\rho} = \xi_{\mu}$ should be excluded because the 
corresponding solution to the Lippmann-Schwinger equation is not normalizable.
In addition, it must be required that
$\xi_{\rho}>\xi_{\mu}$ so that the width is strictly non-negative. This condition is fulfilled 
for all material pairs up to our knowledge (as long as $\xi_{\rho}>0$ as stated earlier) \cite{materials},
\cite{materials2}. 

Finally,
plugging the residue matrix into Eq. (\ref{eq:res}) the poles of the reflection
matrix for the composite structure can be determined.
 The widths of these poles are given by
$$
\Gamma_{\pm} = \tilde{\Gamma}\Big[\Big(1\mp\cos({p_{l,x}d})\Big)w_{l}^{2}+
\Big(1\mp\cos(p_{t,x}d)\Big)w_{t}^{2}\Big],
$$
where the sign
$\pm$ corresponds to even and odd parity as before. 
One can see that the condition for the existence of a BSC is given by
\begin{equation}\label{eq:SW}
p_{l,x}d=\pi M<p_{t,x}d=\pi N,
\end{equation}
where $N$ and $M$ are either mutually even integers for even parity 
or mutually odd integers for odd parity. 
It should be noted that when the even/odd parity state turns into a BSC the odd/even parity resonances has double 
the width much like in the electromagnetic
case \cite{PRL2008}. 

Elastic BSCs are robust 
under variations of elasticity parameters. The condition (\ref{eq:SW}) guarantees that one can construct a standing waves (BSC) composed of two polarizations even though these polarization modes are coupled. For the frequency range under consideration (two open diffraction channels),
the results do not depend on the longitudinal Lam\'e coefficient, $\xi_{\lambda}$. The effect of this parameter can only be seen near longitudinal diffraction thresholds which requires an analysis of higher open diffraction thresholds. 
An analysis of multiple open channels and intermediate distances is very technical. However special cases will be analyzed in Section 4.

\subsection{Numerical Analysis of BSC in the Fabry-Perrot Limit} 

The perturbation theory developed in the previous section showed that the scattering of the in-plane modes
is resonance-dominated, and the background scattering can be neglected in the leading order.
Here the problem is investigated numerically without using the perturbation theory. The objective 
is show that the results of the perturbation theory agrees numerical studies of the exact theory. 
The case $\xi_{\mu}=0$ is considered because of its simplicity and that
this parameter plays a similar role to $\xi_{\rho}$ and 
does not affect anything relevant for our objective and in physics of the system (as will be shown below).

For spatially homogeneous Lam\'e coefficients, (\ref{2.2.5}) and (\ref{2.2.6}) are reduced to 
$$
\bar{u}_{i}=S_i(\omega^2)\bar{u}_{i}^{0}\,,\quad S_{i}(\omega^2)=\frac 1{1+\omega^{2}\xi_{\rho}\Pi_{ii}(0)}\,.
$$
The exact reflection matrix is obtained  in the same way as in the previous section
\begin{eqnarray*}
u^{R}_{t}&=& R_{tl}(\omega^{2},k_{y})u_{l}^{0}+R_{tt}(\omega^{2},k_{y})u_{t}^{0}\,,\\
u^{R}_{l}&=& R_{ll}(\omega^{2},k_{y})u_{l}^{0}+R_{lt}(\omega^{2},k_{y})u_{t}^{0}\,,
\end{eqnarray*}
where the matrix elements read
\begin{eqnarray*}
R_{ll}(\omega^{2},k_{y})&=&\frac{i\epsilon \xi_{\rho}}{2k_{l,x}}\Big(k_{y}^{2}S_y(\omega^2)-k_{l,x}^{2}
S_x(\omega^2)\Big)\,,\\
R_{lt}(\omega^{2},k_{y})&=&\frac{i\epsilon\xi_{\rho}\alpha }{2k_{l,x}}
\Big(k_{t,x}k_{y}S_y(\omega^2)+k_{l,x}k_{y}S_x(\omega^2)\Big)\,,\\
R_{tt}(\omega^{2},k_{y})&=&\frac{i\epsilon\xi_{\rho}}{2k_{t,x}}\Big(k_{t,x}^{2}S_y(\omega^2)-
k_{y}^{2}S_x(\omega^2)\Big)\,,\\
R_{tl}(\omega^{2},k_{y})&=&\frac{i\epsilon\xi_{\rho} }{2\alpha k_{t,x}}\Big(k_{t,x}k_{y}S_y(\omega^2)+
k_{l,x}k_{y}S_y(\omega^2)\Big)\,.
\end{eqnarray*}
%\textcolor{red}{(THIS IS WRONG, DELETE) - It is worth noting that for homogeneous Lam\'e coefficients the longitudinal polarization is parallel to the 
%${x}$ axis while the in-plane transverse polarization is parallel to the ${y}$ axis.}
%
%\textcolor{red}{(MAYBE YOU MEANT THIS - It is worth noting that for $k_{y} = 0$ the longitudinal polarization is parallel to the 
%${x}$ axis while the in-plane transverse polarization is parallel to the ${y}$ axis.}

Let us show that terms proportional to $S_x(\omega^2)$ describe the resonance scattering, while those 
proportional to $S_y(\omega^2)$ contribute only to the background scattering, by investigating
the values of $S_{x,y}(\omega^2)$ along
the parameter curves given in (\ref{tau}) and (\ref{Delta_-1}).
Using the series representations for $\Pi_{yy}(0)$ given in Appendix A3, 
\begin{eqnarray}\label{3.4.1}
1+\omega^{2}\xi_{\rho}\Pi_{yy}(0) &=& 1+O(\epsilon \ln{\epsilon})\,,\\ \label{3.4.2}
1+\omega^{2}\xi_{\rho}\Pi_{xx}(0) &=& 1-\frac{\tau}{\tau_{0}}+\epsilon \Sigma(\epsilon)
\end{eqnarray}
where $\tau_{0} =2c_t (\omega_{t,-1}^{2} \xi_{\rho})^{-1}$ and
$\Sigma(\epsilon)$ is analytic in $\epsilon>0$ and $\epsilon\Sigma(\epsilon)\to 0$
as $\epsilon\to 0^+$. Its explicit form
can be deduced using the Schl\"omilch
series for $\Pi_{xx}(0)$ given in Appendix A3. Therefore
\begin{eqnarray}\nonumber
\epsilon S_y(\omega^2)&=& O(\epsilon)\,,\\ \label{Sx}
\epsilon S_x(\omega^2) &=&
\frac{\zeta}{\omega^{2}-\omega_{t,-1}^{2}+\Delta_{0}^{2}}\,,\\ \nonumber
\Delta_{0} &=& \frac{\epsilon}{\tau_{0}(1+\epsilon \Sigma(\epsilon))}\,,\\ \nonumber
\zeta &=& -\frac{\epsilon \Delta(\Delta+\Delta_{0})}{1+\epsilon \Sigma(\epsilon)} = -\frac{2 \epsilon^{3}}{\tau_{0}^{2}}+O(\epsilon^{4}\log{\epsilon})\,,
\end{eqnarray}
The pole $\omega_0^2-i\Gamma$ of $S_x(\omega^2)$ is obtained from the equation 
$$
\omega_{0}^{2}-i\Gamma=\omega_{t,-1}^{2}-\Delta_{0}^{2}= \omega_{t,-1}^{2}(1-\frac{1}{4}\epsilon^{2}\xi_{\rho}^{2}k_{y,-1}^{2})+\frac{2\epsilon^{3}}{\tau_{0}^{2}}\Sigma(\epsilon) +O(\epsilon^4)
$$

In the perturbation theory in small $\epsilon$, 
the width $\Gamma$ is determined by the imaginary part of $\Sigma(\epsilon)$ in the leading order. Using 
the Schl\"omilch series for $\Pi_{xx}(0)$, it is inferred that 
\begin{eqnarray}\nonumber
\Sigma(\epsilon) &=& \Sigma_0(\epsilon)+O(\epsilon^{2}\log{\epsilon})\,,\\ \label{Sigma0}
\Sigma_0(\epsilon) &=& -\frac{i \xi_{\rho}}{2}\Big\{\sum_{n \neq 0}\Big(\frac{\Omega_{l,n}}{c_l}+
\frac{i\omega_{t,-1}^{2}}{4\pi c_{l}^{2} |n|}\Big)+\sum_{n \neq 0, -1}\Big(\frac{\omega_{t,n}^{2}}{c_t\Omega_{t,n}}+\frac{i\omega_{t,-1}^{2}}{4\pi c_{t}^{2} |n|}\Big) \\ \nonumber
&&+\frac{i\omega_{t,-1}^{2}}{4\pi c_{t}^{2} }+\frac{\Omega_{l,0}}{c_{l}}+
\frac{\omega_{t,0}^{2}}{c_t\Omega_{t,0}}-\frac{i \omega_{t,-1}^{2}}{2 \pi c_{t}^{2}}+\frac{i \omega_{t,-1}^{2}(1+\alpha^{2})}{4 \pi c_{t}^{2}}\log(4 \pi \epsilon) \Big \}\,,
\end{eqnarray}
where $\Omega_{t,n}=(\omega_{t,-1}^{2}-\omega_{t,n}^{2})^{1/2}$ and 
$\Omega_{l,n}=(\omega_{t,-1}^{2}-\omega_{l,n}^{2})^{1/2}$. Therefore

\begin{eqnarray}\label{in_plane_pole}
\omega_{0}^{2} &=& \omega_{t,-1}^{2}(1-\textstyle{\frac{1}{4}}\epsilon^{2}\xi_{\rho}^{2}k_{y,-1}^{2})+
O(\epsilon^{3}\ln{\epsilon})\,,\\ 
\Gamma &=& \frac{\epsilon^{3}\xi_{\rho}^{3}\omega_{t,-1}^{4}}{4c_{t}^{2}} (p_{l,x}+k_{y}^2p_{t,x}^{-1})+O(\epsilon^{4}\ln{\epsilon})\,.
\end{eqnarray}

Next, the perturbation theory is compared with the exact result computed numerically. The exact expression 
for the resonance factor $|S_x(\omega^2)|^2$ is computed and plotted in Figure 2 (solid black line)
for parameters specified in the caption. The obtained
profile is numerically fit into a standard Lorentzian profile where 
$S_x=i\Gamma(\omega^2-\omega_0^2+i\Gamma)^{-1}$
with the position of the pole $\omega^2_0$ and its width $\Gamma$
being the fitting parameters (dashed red line). As one can see, the exact resonance factor in the scattering 
matrix is almost indistinguishable from the standard Lorentzian profile. So, the scattering is indeed 
resonance-dominated. The blue dotted line shows the Lorentzian profile with the position of the pole
$\omega_0^2$ 
and its width $\Gamma$ calculated by the perturbation theory in the leading order in $\epsilon$ for the same values
of the parameters. The relative error of the perturbation theory in the pole position is
$1.1\times 10^{-5}\%$ and in the width it is $1.461\%$ for the material and geometrical parameters specified in the plot. An analysis of the Fabry Perrot limit of the in-plane polarizations for a periodic double array of elastic scatters was first presented in \cite{Wave_Motion}, it will be shown in the following sections that these BSC continue to persist for arbitrary values of $\omega d$. It's interesting to note that the phase matching condition (\ref{eq:SW}) turns out to be exact, while the resonances frequency is modified for intermediary separations, $d$.
\begin{figure}[!htb]
	\begin{center}
		\minipage{0.4\textwidth}
		% Lorentz Fit
		\includegraphics[width=3.5in,height=2in]{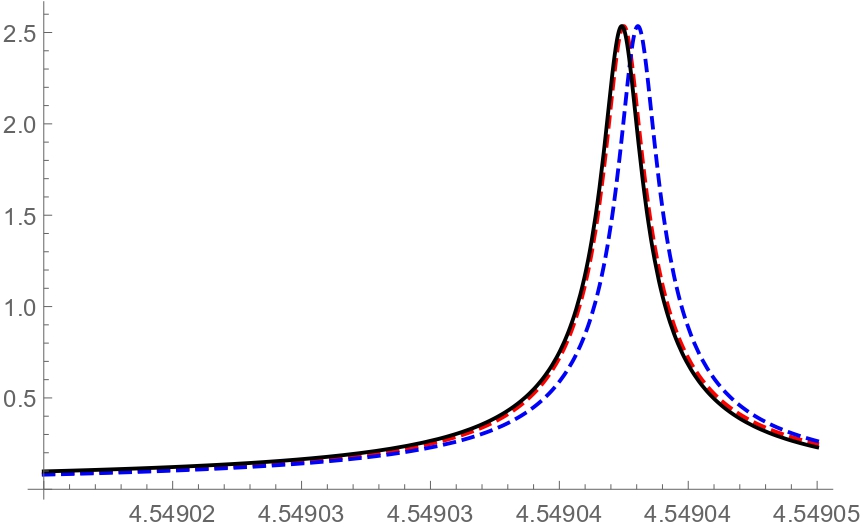}  
		\endminipage
	\end{center}
	\caption{\small Plot of $|\epsilon S_x(\omega^2)|^2$ vs $q$, where $\omega=c_{t}q$ for the following 
parameters $(k_{y},\epsilon, d,\xi_{\rho},\alpha)=(1.73405,.001,29880,.280,.59915)$
(black solid line). The red dashed line correspond to a numerical Lorentzian fit with fitting parameters given by $(\omega_{0},\Gamma)=(4.5490425c_{t},6.72071 \cdot 10^{-6}c_{t}^{2})$. The blue dotted line corresponds to the theoretical fit from the expansion outlined below with $(\omega_{0},\Gamma)=(4.5490430c_{t},6.62285 \cdot 10^{-6}c_{t}^{2})$}
\end{figure}

\section{Asymmetric Double Array}
 
Consider two periodic identical arrays at a distance $d$ such that one array is shifted parallel relative to 
the other by a distance $0\leq \delta <1$ so that the centers of the cylinders are at 
\begin{eqnarray*}
r_{n,-} &=& -\frac{d}{2}\hat{x}+n\hat{y} \,,\\
r_{n,+} &=& \frac{d}{2}\hat{x}+(n+\delta)\hat{y}\,,
\end{eqnarray*}
with $n$ being an integer. The objective is to investigate the existence of BSC in this system. In addition,
the number of open diffraction channels is not limited in contrast to the preceding discussion, and the analysis
of poles of the scattering matrix is carried out without the Fabry-Perrot approximation. 
Unfortunately, 
an analysis for general Lam\'e coefficients
cannot be done without a substantial numerical assistance and will not be presented 
here. However, a special case $\xi_\mu=\xi_\lambda=0$
can be studied analytically. As explained above, $\xi_\rho >0$.
A further technical assumption is that the lowest closed diffraction
channel is transverse. Its threshold is denoted by $\omega_{t,N}^2$ for some integer $N$. Similarly,
the threshold for the lowest closed longitudinal diffraction channel is given by $\omega^2_{l,M}$
for some integer $M$, and $\omega_{t,N}^2<\omega^2_{l,M}$. As before, the scattering of in-plane 
and out-of-plane polarization modes is decoupled. For the out-of-plane mode, the problem is fully analogous
to the electromagnetic counterpart studied in \cite{JMP2010}. So, here only BSC for the in-plane modes 
are investigated.

\subsection{BSCs for the in-plane modes}

BSC consisting of mixed (in-plane) polarization correspond to non-trivial solutions to the associate homogeneous  equations
(\ref{2.2.10}) and (\ref{2.2.11}) for $\omega^2>\omega_{l,0}^{2}$, in an open diffraction channel. 
With $\xi_\lambda=\xi_\mu=0$, Eq. (\ref{2.2.11}) is identically satisfied (since the tensors 
$K^{(1,2)}$ vanish), and
the homogeneous equation (\ref{2.2.10}) is simplified to 
\begin{equation}\label{4.1}
(1+\omega^{2}\xi_{\rho}\ \Pi_{ii}(0))\bar{u}^{\pm}_{i}+\omega^{2}\xi_{\rho}\bar{u}^{\mp}_{j}\Pi_{ji}(\pm \mu_{0})= 0,
\end{equation}
where $\mu_{0}=d\hat{x}+\delta\hat{y}$. The equation is decoupled for $\bar{u}_z^\pm$ and $\bar{u}^\pm_{x,y}$
as already noted. In what follows, the indices ranges over $x$ and $y$ (in-plane) components. Equation (\ref{4.1})
can be reduced to two separate equations for $\bar{u}_i^+$ and $\bar{u}_i^-$
\begin{eqnarray*}
&&(I-S^{\pm})_{il}\bar{u}_{l}^{\pm}=0\,,\\
&&S^{\pm}_{il}=\sum _j\frac{\omega^{4}\xi_{\rho}^{2}\tilde{\Pi}^{\mp}_{lj}\tilde{\Pi}^{\pm}_{ji}}{(1+\omega^{2}\xi_{\rho}\Pi_{ii}(0))(1+\omega^{2}\xi_{\rho}\Pi_{jj}(0))}.
\end{eqnarray*}
where $\tilde{\Pi}^{\pm}_{ji}=\Pi_{ji}(\pm \mu_{0})$ for brevity. A non-trivial solution exists if and only if
$\det(I-S^{\pm})=0$. It follows from (\ref{3.4.1}) and (\ref{3.4.2}) that the latter equation is reduced 
in the leading order to $1-S_{xx}=0$ or
\begin{eqnarray}\label{eq:TransEq}
(1+\omega^{2}\xi_{\rho}\Pi_{xx}(0))^{2}= \omega^{4}\xi_{\rho}^{2}\tilde{\Pi}_{xx}^{+}\tilde{\Pi}_{xx}^{-},
\end{eqnarray}
the corresponding non-trivial solution reads
$$
(\bar{u}_{x}^{+})^2=(\bar{u}_{x}^{-})^{2}\neq 0\,,\quad \bar{u}_{y}^{\pm}=0\,.
$$
The amplitude $|\bar{u}_x^\pm|$ is included into a normalization constant of the corresponding 
solution (Seigert state) to the Lippmann-Schwinger equation.

A BSC exists if
(\ref{eq:TransEq}) has a positive real root that lies
in open diffraction channel , $\omega^{2} > \omega_{t,0}^{2}$, where $\omega_{t,0}^2$ is the continuum edge (the lowest frequency
squared for all propagating modes in the asymptotic region $|x|\to \infty$ for a given $k_y$).
As stated earlier, we consider the case when the lowest closed diffraction threshold is transverse and is given by $\omega_{t,N}^{2}$, we define  $\Delta^{2} = \omega_{t,N}^{2}-\omega^{2}$. As before, solutions to (\ref{eq:TransEq}) are sought
along a parametric curve (\ref{tau}).
The real and imaginary parts of the left- and right-hand sides of (\ref{eq:TransEq}) 
are to be expanded 
to leading order along this curve. In the left hand-side, the expansions (\ref{3.4.2}) and (\ref{Sigma0}) are 
used where $\tau_{0} = \omega_{t,N}^{2} \xi_{\rho}/2c_{t}$ and
$\Sigma_0$ has the same form as in (\ref{Sigma0}) in which the diffraction threshold
$\omega_{t,-1}^2$ is replaced by $\omega_{t,N}^2$ and the summation index in the second series 
cannot take value $N$. In the right-hand side, by using the Scl\"omilch series for $\Pi_{xx}(r)$ given 
in Appendix A3, it is found that
\begin{eqnarray*}
\omega^{2}\xi_{\rho}\tilde{\Pi}_{xx}^{\pm} &=&-(\tau/\tau_{0})e^{-\frac{\Delta d}{c_{t}}}e^{\pm ik_{y,N}\delta}-\epsilon g(d,\pm \delta)+O(\epsilon^{3})\,,\\
g(d,\pm \delta) &=& -\frac{i \epsilon \xi_{\rho}}{2} \sum_{n \neq N} \Big[\frac{\Omega_{l,n}}{c_{l}} \
{\phi}_{n}(k_{l,N},k_{y};\pm \mu_{0})+\frac{1}{c_{t}} \frac{\omega_{t,n}^{2}}{\Omega_{t,n}}\
{\phi}_{n}(k_{t,N},k_{y},\pm \mu_{0})\Big]\,,\\  
{\phi}_{n}(k,k_{y};r)&=& e^{i(k_{x,n}|x| +k_{y,n}y)}\,,
\end{eqnarray*}

where $k_{y,n}$ is defined in (\ref{k_yn}), $\Omega_{a,n}=(\omega_{t,N}^{2}-\omega_{a,n}^{2})^{1/2}$, 
$k_{x,n}=(k^2-k_{y,n}^2)^{1/2}$, and $c_{a}k_{a,N} = \omega_{t,N}$.
To leading order the real part of (\ref{eq:TransEq}) is given by 
\begin{eqnarray}\label{eq:RlTransEq}
\Big(1-\frac{\tau}{\tau_{0}}\Big)^{2} = \Big(\frac{\tau}{\tau_{0}}\Big)^{2}e^{-\frac{2 \Delta}{c_{t}} d}\,.
\end{eqnarray}
Taking the square root of both sides, two equations are obtained
\begin{eqnarray*}
1-\frac{\tau^{\pm}}{\tau_{0}} = \pm \Big(\frac{\tau^{\pm}}{\tau_{0}}\Big)e^{-\frac{ \Delta^{\pm}}{c_{t}} d}\,,
\end{eqnarray*}
where $\Delta ^\pm =\epsilon/\tau^\pm$. The parity index
$\pm$ corresponds to even/odd states as was discussed in the prior sections. By examining the graphs 
of functions $(1-q)/q$ and $\pm e^{-a/q}$, where $q=\tau/\tau_0$ and $a>0$,
it is not difficult to see that each of these transcendental equations has just one solution if $q>0$.
They will be discussed in the next section and obtained in the case 
of two open diffraction channels. 

If $\tau$ solves (\ref{eq:RlTransEq}), 
the leading order for the imaginary part of (\ref{eq:TransEq}) is given by
\begin{eqnarray}\label{eq:ImTransEq}
\mathrm{Im}(e^{ik_{y,N}\delta}g(d,-\delta)+e^{-ik_{y,N}\delta}g(d,\delta)) = \pm 2 \mathrm{Im}(g(0,0)).
\end{eqnarray}
Using the explicit form of the function $g$, this equation can be reduced to
\begin{eqnarray*}
0&=& \sum_{n \in N_{l}^{(o)}} A_{n}+\sum_{n \in N_{t}^{(o)}} B_{n}\,,\\
A_n&=&
\frac{\Omega_{l,n} }{c_l} \Big[1 \mp \cos(2 \pi (N-n) \delta)
\cos\Big(\frac{\Omega_{l,n} d}{c_l}\Big)\Big]\,,\\
B_n&=&
\frac{\omega_{t,n}^{2}}{c_t\Omega_{t,n}}\Big[1 \mp\cos(2 \pi (N-n) \delta)
\cos\Big(\frac{\Omega_{t,n} d}{c_t}\Big)\Big]\,,
\end{eqnarray*}

where $N_a^{(o)}$ denotes the range of $n$ corresponding to open diffraction channels for the 
polarization mode $a=l,t$.
Since each term in the series is non-negative, the equation is satisfied 
only if $A_n=B_n=0$. This can be possible only if $\cos(2\pi\delta)=\pm 1$
or $\delta = 0, \frac{1}{2}$ because $|\cos(\frac{\Omega_{a,n}d}{c_{a}})|\leq 1$.
 It is then concluded that
in-plane BSCs for a shifted double array do not exist if $0<\delta<1$ and $\delta\neq \frac 12$.
It is noteworthy that the conclusion does not involve any approximations.

%\textcolor{green}{It is unclear what the following blue text is meant for. Either clarify or remove}
%\textcolor{yellow}{
%As an example we can take the positive parity states with $\delta = 0$, if we consider the case where $\alpha$ and $k_{y}$ are fixed then we must check if there is a value of $d$ such that the above equation can be satisfied, define the set
%\begin{eqnarray*}
%D_{a,n} = \{\frac{2 \pi c_{a} m}{\sqrt{\omega_{t,N}^{2}-\omega_{a,n}^{2}}} : m \in N\}.
%\end{eqnarray*}
%Define the following set
%\begin{equation*}
%D_{0} = \bigcap_{a \in \{t,l\}}\bigcap_{n \in N_{a}^{c}}D_{a,n}.
%\end{equation*}
%Clearly a solution to the above equation exist if and only if $D_{0}$ is not empty, if $D_{0}$ is non-empty then any $d$, such that $d \in D_{0}$ will satisfy the above equation. The situation is slightly more complicated if $\delta = \frac{1}{2}$, in this case we must define 
%\begin{eqnarray*}
%D_{a,n}^{+} = \{\frac{2 \pi c_{a} m}{\sqrt{\omega_{t,N}^{2}-\omega_{a,n}^{2}}} : m \in N\},
%\end{eqnarray*}
%if $N-n$ is even or zero, otherwise if $N-n$ is odd we must define
%\begin{eqnarray*}
%D_{a,n}^{-} = \{\frac{(2 m +1)\pi c_{a} }{\sqrt{\omega_{t,N}^{2}-\omega_{a,n}^{2}}} : m \in N\}.
%\end{eqnarray*}
%Define
%\begin{equation*}
%D^{+} = \bigcap_{N-n = 0, 2, min}
%\end{equation*}
%}
%\subsection{Mixed Polarization BSC}
    
In particular, consider the case $N=-1$ and the spectral range specified by (\ref{eq:spectral range}).
Then Eq. (\ref{eq:ImTransEq}) is reduced to
$A_0=0$ and $B_0=0$, which can be written in the form 
\begin{eqnarray*}
1 \mp \cos(2 \pi \delta)\cos(p_{l,x}d)&=&0\,,\\
1 \mp\cos(2 \pi \delta)\cos(p_{t,x}d) &=& 0
\end{eqnarray*}
where $p_{a,x}=\Omega_{a,0}/c_a$ for $N=-1$ (they were introduced earlier in Section 3.3).
When there is no offset, $\delta=0$, these equations are nothing but the phase matching condition given in (\ref{eq:SW}). For $\delta = \frac{1}{2}$ the phase matching conditions for the even and odd parity states are switched, while the resonance frequency is kept the same. In any case this phase fixing is identical to the result that was found in the Fabry-Perrot limit \cite{Wave_Motion} and can clearly be satisfied by fixing the phase in the matter prescribed in the previous section, namely, (\ref{eq:SW}). Thus, the analysis
shows that, first, the round-trip phase matching condition is determined purely by the far field propagating modes,
and, second, once it is fulfilled, the corresponding resonance state becomes a BSC whose frequency
is determined by (\ref{eq:RlTransEq})  that always has a real solution. Comparing this conclusion to the Fabry-Perrot
approximation analysis, it is noteworthy that the coupling of resonances in two arrays via  
evanescent modes only affects the frequency of a BSC, whereas the phase matching condition does not depend on it.  

%\textcolor{blue}{(\ref{eq:newpoles})}

\subsection{Explicit form of BSCs}
Let us analyze (\ref{eq:RlTransEq}) , it should be noted that resonances frequency is independent of the shift, $\delta$, to the leading order. The equation can be rearranged to
a more tractable form
$$
\frac{\Delta^{\pm}}{1\mp e^{-\frac{\Delta^{\pm}d}{c_{t}}}}=\frac{\epsilon \xi_{\rho}\omega_{t,N}^{2}}{2c_{t}}=\tilde{\Delta}\,.
$$
Note that only the right-hand side of this equation depends on the number of open transverse diffraction channels through the constant $\tilde{\Delta}$. Let $\Delta^{\pm}=c_{t}s^{\pm}$ and $\tilde{\Delta}=c_{t}\tilde{s}$, then 
a solution to the above equation can be expressed in terms of product log functions, $W_{n}(z)$:
$$
\frac{s^{\pm}}{\tilde{s}}=\frac{W_{0}(\mp \tilde{s}d e^{-\tilde{s}d})}{\tilde{s}d}+1,
$$
where 
$$
\ln{W_{n}(z)}=\ln{z}-W_{n}(z)+2\pi i n.
$$ 

\begin{figure}[!htb]
\begin{center}
\minipage{0.4\textwidth}
% Base BSC frequency
\includegraphics[width=3.5in,height=2in]{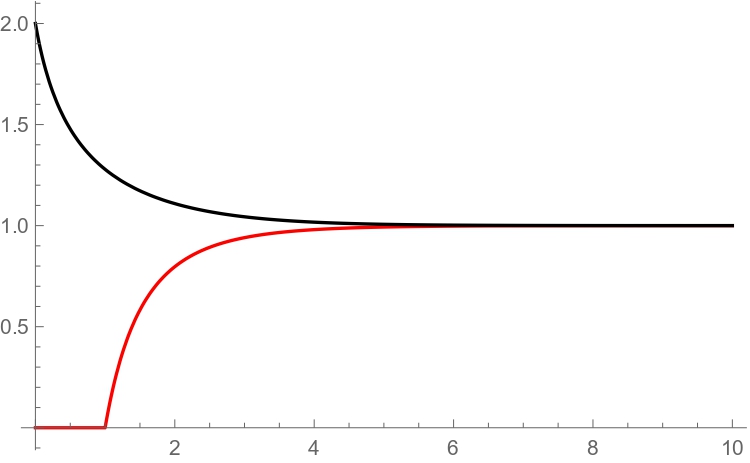} \endminipage
\end{center}
\caption{\small Plot of $\frac{s^{\pm}}{\tilde{s}}$ vs $\tilde{s}d$. The red curve corresponds to the even states, $(\frac{s^{+}}{\bar{s}})$, and the black curve corresponds to the odd states, $(\frac{s^{-}}{\bar{s}})$.}
\end{figure}
\noindent
Note that for $\tilde{s}(2R)<\tilde{s} d<1$ there is no even parity BSC, otherwise there always exist two BSCs for sufficiently small scatters, whose frequency is given by

$$
\omega_{BSC,\pm}^{2}= \omega_{t,N}^{2}-(c_{t}s^{\pm})^{2}.
$$

The corresponding field is given by
$$
\frac{u_{i}^{\pm}(r)}{\bar{u}_{x}^{\pm}}=-\omega_{BSC,\pm}^{2}\xi_{\rho}[\Pi_{xi}(r-{\textstyle \frac{d}{2}}\hat{x})
\mp\Pi_{xi}(r+{\textstyle \frac{d}{2}}\hat{x})].
$$

For the $(+)$ sign, $x$-component of the field is odd with respect to $x$ and the $y$-component of the field is is even, the case of $(-)$ corresponds to an even $x$-component of the field and an odd $y$-component of the field with respect to $x$, this is due to the parity symmetry associated with reflections about the $\hat{y}$ axis. With  regard to the full parity transformation, the vector $u^{\pm}(r)$ is even for the $(+)$ and odd for $(-)$. In either case, in the asymptotic region $|x| \rightarrow +\infty$,
$$
u_{i}^{\pm}(r)\sim e^{-s^{\pm}\mid x\mid}e^{i(k_{y}+2\pi N)y}
$$

up to a phase. It is clear that this solution is square integrable and satisfies all the boundary conditions listed in the introduction, hence, it is a BSC consisting of two polarization modes with different group velocities. 
This analysis agrees with the Fabry-Perrot limit results considered in the previous section for $N=-1$, because, as $\Delta d \rightarrow +\infty$, $s^{\pm} \rightarrow \tilde{s}$, so that
$$
\omega_{BSC,\pm}^{2} \to \omega_{t,-1}^{2}-(c_{t}\tilde{s})^{2}=\omega_{0}^{2},
$$
where $\omega_{0}^{2}$ is given in (\ref{in_plane_pole}). In this limit, the parity eigenstates become degenerate to leading order as shown in Figure 3. The case $\xi_{\lambda},\xi_{\mu} \neq 0$ is considerably more difficult, at least from an analytic perspective. 
However, in view of the previous results in Section 3.3 one can surmise that these BSCs will persist through a continuity argument, however a full analysis will require a substantial computational aid and will not be given here.

\subsection{Normal incidence}
It turns out that it is impossible to form a BSC of mixed polarizations in the spectral range under consideration 
if $k_{y}$ is below the lower bound given in (\ref{eq:kybound}); this is due to the symmetry of the array structure, the $\hat{x}$ and $\hat{y}$ components of the field becomes the longitudinal and transverse mode respectively when $k_{y}=0$.
In this case the transverse and longitudinal modes are decoupled, and the transverse transmission coefficient becomes unity to leading order, $t_{t}(\omega)\sim 1+O(\epsilon)$. Any transverse wave would pass 
through the structure with negligible reflection. Therefore
it would be impossible to confine a wave polarized along the $\hat{y}$. However,
longitudinal BSCs still exists in this case. An explicit construction of these single-mode BSCs is 
analogous to that given in \cite{JMP2010} and, for this reason, is omitted here.

\section{\Large Conclusions}

BSC can be supported in a periodic double array consisting of small elastic scatterers and narrow resonances are seen to exist in the single array. Owing to the normal traction boundary conditions, the two in-plane modes (longitudinal and transverse) are decoupled from the out of plane transverse mode.
BSCs were proved to exist for the in-plane and out-of-plane modes. The former
BSC are formed by standing waves in two polarization states that have different dispersion relations
and are coupled through the boundary conditions.
For this reason, they
are significantly more complex then the ones present in similar acoustic and photonic structures. 
Exact analytical solutions for these BSCs are constructed and compared to a partial wave summation in the Fabry-Perrot limit, agreement between the exact solution and the partial wave summation are confirmed, explicitly through an agreement on the round-trip phase condition. 

Further analysis is conducted on the existence of BSC in the asymmetric double array and in higher diffraction channel, it is shown that BSC can exist in higher diffraction channels for certain offsets as long as the round-trip phase matching condition can be met. For the case of two open channels (one transverse and one longitudinal) the exact round-trip phase matching condition can be met by varying the distance between the array as well as either the ratio between the group velocities and/or the Bloch phase. For the single array it is shown that the out of plane transverse mode can support narrow resonances that differ significantly from the those in photonic arrays, this results can be understood in terms of the competing effects between variation in density and variation in Lam\'e coefficients, as a result the width admits a global minimum in parameter space, one that cannot be achieved in photonics and acoustic structures. It has been shown by tuning the variation in density and variation in Lam\'e coefficients one can reduced the width by at least another order of magnitude in the cross section. 

Such BSC and narrow resonances can be used as elastic wave guides or as resonators 
with high quality factors  in a broad spectral
range, especially in view of the fact that elastic systems supporting BSC 
can be designed using mechanical metamaterials (as materials with 
desired elastic properties). In particular, owing to a high sensitivity 
of the quality factor to geometrical and physical properties of 
a resonating system, elastic BSC can be used to detect 
impurities in solids from variations of the density. The energy density 
of a high quality resonance (near-BSC state) has a ``hot'' spots where
it exceeds the energy density of the incident wave by orders in magnitudes,
which would facilitates studies of  non-linear effects in solids.   

\section*{\Large Appendix A}
\subsection*{A1. Green's function}
The needed Green's function is a 
fundamental solution for the differential operator 
$D_{ij}(\nabla)\equiv (\omega^{2}+c_{t}^{2}\triangle)\delta_{ij}+(c_{l}^{2}-c_{t}^{2})\nabla_{i}\nabla_{j}$,
$$
D_{ij}(\nabla)G_{jn}(r)=\delta(r)\delta_{in}
$$
that satisfies the Sommerfeld radiation boundary condition.
The problem is solved by taking the Fourier transform of the equation:
$$
D_{ij}(-ik) F\{G_{jn}\}(k)=\delta_{in}
$$
Since $F\{G_{jn}\}$ is an $SO(2)$ symmetric $2-$tensor, its most general form is
$$
F\{G_{jn}\}(k)= \delta_{ij}g_1(k)+k_jk_ng_2(k)
$$
where $g_{1,2}$ are tempered distributions that satisfy the scalar equations
\begin{eqnarray*}
\Lambda_t(k)g_1(k)&=&1\,,\\
\Lambda_l(k)g_2(k)&=&(c_l^2-c_t^2)g_1(k)
\end{eqnarray*}
where $\Lambda_{a}(k)=\omega^{2}-c_{a}^{2}k^{2}$ with $a=t,l$ labeling the transverse (shear)
and longitudinal (compression) modes. Since $\Lambda_a$ are polynomials, any solution to the first
equation can written in the form
$$
g_1(k)={\rm Reg}\Big[\frac{1}{\Lambda_{t}(k)}\Big]
$$
where the symbol ${\rm Reg}$ stands for a prescription for regularizing the pole
at $|k|=\omega/c_a$ in the $k$ plane so that $g_1$ is a tempered distribution.
A regularization is needed because the reciprocal of $\Lambda_a$ is not locally  
integrable in the $k$ plane. The regularization is not unique but it is proved 
to exist for any polynomial. Then any solution to the second equation can be written 
in the form
$$
g_2(k)=\frac{c_l^2}{\omega^2}\,{\rm Reg}\Big[\frac{1}{\Lambda_{l}(k)}\Big]-
\frac{c_t^2}{\omega^2}\,{\rm Reg}\Big[\frac{1}{\Lambda_{t}(k)}\Big]
$$
Its verification is based on the distributional equality
$$
\Lambda_l(k)\,{\rm Reg}\Big[\frac{1}{\Lambda_{t}(k)}\Big]=
\frac{c_l^2}{c_t^2}+\frac{(c_t^2-c_l^2)\omega^2}{c_t^2}\,{\rm Reg}\Big[\frac{1}{\Lambda_{t}(k)}\Big]
$$
which, in turn, follows from the identity
$c_l^2\Lambda_t-c_l^2\Lambda_l=(c_l^2-c_t^2)\omega^2$.

 The regularization should be chosen so that 
the inverse Fourier transform of $F\{G_{ij}\}$ satisfies the Sommerfeld condition. This is achieved by using 
the $+i0^+$ prescription to shift the pole so that 
$$
{\rm Reg}\Big[\frac{1}{\Lambda_{t}(k)}\Big]=\frac{1}{\Lambda_a(k)+i0^+}
$$
The inverse Fourier transform of the above distributions reads
\begin{equation}\label{FH0}
F^{-1}\Big\{\frac{1}{\Lambda_a(k)+i0^{+}}\Big\}(r)=-\frac{i}{4c_a^2}H_{0}^{(1)}(k_{a}|r|)\,,
\end{equation}
where $k_a=\omega/c_a$ and
$H_0^{(1)}$ is the Hankel function of the first kind. Using this equation, 
the final expression for the Green's
function is obtained
$$
G_{jn}(r)=-\frac{i}{4c_{t}^{2}}\Big(\delta_{jn}+\frac{\nabla_{j}\nabla_{n}}{k_{t}^{2}}
\Big)H_{0}^{(1)}(k_{t}|r|)+\frac{i}{4c_{l}^{2}}\Big(\frac{\nabla_{j}\nabla_{n}}{k_{l}^{2}}
\Big)H_{0}^{(1)}(k_{l}|r|)\,.
$$
The Green's function is a regular distribution (a locally integrable function
in a plane) and smooth everywhere but $r=0$. It is readily to see from the asymptotic behavior of the 
Hankel functions that the Green's function satisfies the Sommerfeld (outgoing wave) condition.

\subsection*{A2. Lippmann-Schwinger equation in the long wavelength limit}
Let us investigate the Lippmann-Schwinger equation (\ref{LSE}) for 
a periodic array of cylinders in long wavelength limit (the radius
of cylinders is much smaller than the wavelength of the incident wave).
To this end, let us calculate the convolution in (\ref{LSE}) in this limit.
If $\chi$ is the characteristic function of the support of the relative 
mass density and Lam\'e coefficients, denoted by $\Omega$, then the problem is reduced to
evaluating the convolution
\begin{eqnarray*}
[H_{0}^{(1)}(k_{a}|r|)*(\chi u_{j})](r)&=&\sum_{n}e^{ik_{y}n}U_{j}(r-r_n)\,,\\
U_{j}(r-r_n)&=&
\int_{|r'|<R} H_{0}^{(1)}
(k_{a}|r-r'-r_n|)u_{j}(r')\,d^{2}r'
\end{eqnarray*}
where $r_n=n\hat{y}$, by the Bloch periodicity condition.
For $r \notin \Omega$, the Grafts formula is used to expand the Hankel function
$$
H_{0}^{(1)}(k_{a}\mid r-r' \mid)=\sum_{n}e^{i(\theta-\theta')n}H_{n}^{(1)}(k_{a}|r|)J_{n}(k_{a}| r'|)
$$
and evaluate the integrals
\begin{eqnarray*}
U_j(r-r_n)&=&\sum_{m}V_{j,m}e^{im\theta_{n}}H_{m}^{(1)}(k_{a}|r-r_n|)\\
V_{j,m}&=&\int_{| r'|<R}J_{m}(k_{a}|r'|)u_{j}(r')e^{-im\theta'}d^{2}r'
\end{eqnarray*}
The advantage is that the Bessel functions are analytic so that 
the behavior of the integrals $V_{j,m}$ for small $\omega R$ can be investigated.

In isotropic elastic scattering, the field is expanded into conservative and rotational parts
$$
u_{i}(r)=\nabla_{i}\phi_{l}(r)+\epsilon_{ijz}\nabla_{j}\phi_{t}(r)+\epsilon_{ijk}\epsilon_{kmz}\nabla_{j}\nabla_{m}\phi_{t,z}(r) 
$$
where $\epsilon_{ijk}$ is the totally skew-symmetric tensor, $\epsilon_{xyz}=1$. 
Inside the scatter the potentials satisfy the Helmholtz equation
$$
(\Delta+p_{a}^{2})\phi_{a}(r)=0
$$
where $a$ labels the potentials,  
$p_{a}^{2}=\omega^{2}/\bar{c}_{a}^{2}$ is the wave vector inside the scatter, 
\begin{eqnarray*}
\bar{c}_{l}^{2}&=&(c_{l}^{2}-2c_{t}^{2})\frac{1+\xi_{\lambda}}{1+\xi_{\rho}}+2c_{t}^{2}\,\frac{1+\xi_{\mu}}{1+\xi_{\rho}}\,,\\
\bar{c}_{t}^{2}&=&c_{t}^{2}\,\frac{1+\xi_{\mu}}{1+\xi_{\rho}}\,,
\end{eqnarray*}
and $c_{a}$ is the group velocity in the scatter for the mode $a$, $c_t=c_{t,z}$.
Its regular solution is obtained by separating variables in the polar coordinates 
$$
\phi_{a}(r)=\sum_{n}c_{a,n}J_{n}(p_{a}|r|)e^{in\theta}
$$
The analysis for
the out-of-plane mode is nearly identical to the electromagnetic case studied in \cite{JMP2010}.
In what follows, only the in-plane modes are investigated, that is, $\phi_{t,z}=0$ in the field $u_i$.
 Put
$$
f_{m}(w,v)=\int_0^wJ_{m}(vs)J_{m}(s)\,s\, ds\,.
$$
Then

\begin{eqnarray*}
\int_{|r'|<R}J_{m}(k_{a}|r'|)e^{-im\theta'}\nabla_{x'}\phi_{b}(r')d^{2}r'
&=&\frac{\pi}{p_{b}}(c_{b,m+1}- c_{b,m-1})f_{m
}\Big(p_{b}R,\frac{k_{a}}{p_{b}}\Big)\,,\\
\int_{|r'|<R}J_{m}(k_{a}| r'|)e^{-im\theta'}\nabla_{y'}\phi_{b}(r')d^{2}r'
&=&\frac{i\pi}{p_{b}}(c_{b,m+1}+c_{b,m-1})f_{m}\Big(p_{b}R,\frac{k_{a}}{p_{b}}\Big)\,.
\end{eqnarray*}

Where $a, b \in \{l, t, z\}$ are polarization subscript. In the limit of small $\omega R$
$$
f_{m}\Big(p_{b}R,\frac{k_{a}}{p_{b}}\Big)= O((\omega R)^{2|m| +2
})
$$
To complete the estimation of the integrals $V_{j,n}$, one should investigate the behavior of
 $c_{a,n}$ when $\omega R\ll 1$. The field $u_j$ has only $x$ and $y$ components.
Define a $2\times 2$ matrix $M_n(\omega R)$ as a linear transformation
of the Fourier trigonometric coefficients of the field $u_j$ at $|r|=R$ and the coefficients
$c_{a,n}$:
$$
\int_{0}^{2\pi}
\begin{pmatrix}u_x(R,\theta)\cr u_y(R,\theta)\end{pmatrix}e^{in\theta}d\theta=M_{n}(\omega R)\begin{pmatrix} c_{l,n}\cr c_{t,n}\end{pmatrix}
$$
By evaluating the integral, it is concluded that
$$
M_{n}(\omega R) =
\left( {\begin{array}{cc}
p_{l}J_{n}'(p_{l}R) & \frac{in}{R}J_{n}(p_{t}R) \\ 
\frac{in}{R}J_{n}(p_{t}R)&-p_{t}J_{n}'(p_{t}R) \\
\end{array} } \right)
$$
Since the field $u_j$ is bounded, the inverse of $M_n$ defines 
the behavior of $c_{a,n}$ in the long wavelength limit:
$$
c_{a,n} =O((\omega R)^{-|n|-1})
$$
This implies that
$$
V_{j,n}=O((\omega R)^{2|n|-|n\pm1|+1})
$$
The lowest order term is at $n=0$ so that
$$
U_j(r-r_n)=H_{0}^{(1)}(k_{a}|r -r_n|)V_{j,0} +O(\omega R)
$$
In this limit, $V_{j,0}=\epsilon \bar{u}_j+O(\epsilon^2)$ where $\bar{u}_j=u_j(0)$ and $\epsilon=\pi R^2$.

A similar analysis can be carried out for the Lam\'e coefficient term in the scattered field so that 
$$
u_{i}(r) = u_{i}^{0}(r)-\epsilon[\omega^{2}\xi_{\rho}\bar{u}_{j}+\bar{\sigma}_{jl}\nabla_{l}]\sum_{n}e^{ik_{y}n}G_{ij}(r-r_n)
$$
in the leading order of the long wavelength approximation. 

\subsection*{A3. Calculation of $\Pi_{ij}(r)$}

In this section, the tensor $\Pi_{ij}$ and its derivatives $(\nabla_n\cdots\nabla_l)\Pi_{ij}=\Pi_{ij, n\cdots l}$ are calculated both
off and on the defects. It follows from the analysis in Appendix A2 that, if $|r-r_n|>R$, 
the integrals in (\ref{PIij}) can be approximated by the integral mean value theorem so that
$$
\Pi_{ij}(r)= \epsilon\sum_{n}e^{ik_{y}n}G_{ij}(r-r_n)
$$
in the leading order in $\epsilon$.
Evaluation of Schl\"omilch series of this type have been discussed in a variety of wave theories \cite{SchlomilchEMseries1}-\cite{SchmomilchSeries15}, \cite{SchmomilchSeries2}-\cite{SchmomilchSeries14}. However an analysis of higher order derivatives requires some special attention since even-order derivatives of Hankel functions are not regular distributions. There are a few methods for evaluating lattice sums. Here a poly-logarithm subtraction method developed in \cite{Rev11}-\cite{Rev12} 
will be invoked
with some modification in order to handle the singular portion of the distributions $\Pi_{ij, n\cdots l}$.

Using the Poisson summation formula in combination with (\ref{FH0}), one infers that \cite{JMP2010}
$$
\frac{1}{2}\sum_{n}e^{ik_{y}n} H_{0}^{(1)}(k|r -r_n|)=\sum_{n}\psi_{n}(k,k_{y};r) 
$$
for $|r-r_{n}|>0$, where $k_{y,n}$ is defined in (\ref{k_yn}), $k_{x,n}=(k^2-k_{y,n}^2)^{1/2}$, and
$$
\psi_{n}(k,k_{y};r)= \frac{e^{i(k_{x,n}|x| +k_{y,n}y)}}{k_{x,n}}\,.
$$
The uniform convergence of $\sum_n\psi_n$ is guaranteed by that $k_{x,n}\sim 2\pi i|n|$ as $|n|
\to \infty$ for all $|x|\geq \delta> 0$ for any positive $\delta$. 
The series converges conditionally if $x=0$ and $y$ is not an interger. 
Therefore in the distributional sense
$$
\frac{1}{2}\sum_{n}e^{ik_{y}n}D_m H_{0}^{(1)}(k| r -r_n|)=\sum_{n}D_m\psi_{n}(k,k_{y};r) 
$$
where for brevity
$$
(D_m)_{i_1i_2\cdots i_m}=\nabla_{i_{1}}\nabla_{i_{2}}\cdots\nabla_{i_{m}}\,.
$$
Since the series $\sum_n D_m\psi_n$ also converges uniformly for $|x|\geq \delta >0$ (by the same 
reason), the asymptotic scattered field can readily be inferred from the expansion
$$
\Pi_{ij}(r)= \frac{i \epsilon}{2c_{l}^{2}} \sum_{n}\Big(\frac{\nabla_{i}\nabla_{j}}{k_{l}^{2}}\Big) 
\psi_{n}(k_{l},k_{y};r)-\frac{i \epsilon}{2c_{t}^{2}} \sum_{n}
\Big(\delta_{ij}+\frac{\nabla_{i}\nabla_{j}}{k_{t}^{2}}\Big) \psi_{n}(k_{t},k_{y};r)
$$
because only open channels with real $k_{x,n}$ contributes in the limit $|x|\to \infty$.

Owing to the Bloch condition, $\Pi_{ij}(r+r_n)=e^{ik_yn}\Pi_{ij}(r)$, it is sufficient
to calculate $D_m\Pi_{ij}(0)$ in order to find the values of $D_m\Pi_{ij}$ at any scatterer:
$$
D_m\Pi_{ij}(0)=\lim_{|r|\to 0^+}D_m\Pi_{ij}(r)
$$
where $\Pi_{ij}(r)$ is defined in (\ref{PIij}). For example
$$
\Pi_{ij}(0)=
 \int_{|r|<R} G_{ij}(r)\,d^2r+\epsilon \sum_{n\neq 0}e^{ik_{y}n} G_{ij}(r_n)\,,
$$
in the leading order of $\epsilon$. Given an explicit form of $G_{ij}$, the problem 
of calculating $\Pi_{ij}(0)$, $\Pi_{ij,l}(0)$, and $\Pi_{ij,nm}(0)$ needed for solving the scattering 
problem, is reduced to evaluating the series of the form
\begin{eqnarray*}
W_m(k,k_{y})&=&\lim_{|r| \rightarrow 0^{+}}w_m(k,k_{y};r)\,,\\
w_m(k,k_y;r)&=&\frac{1}{2}\sum_{n\neq 0 }e^{ik_{y}n} 
D_m H_{0}^{(1)}(k|r- r_n|)
\end{eqnarray*}
for $0\leq m\leq 4$. Note that $W_m$ and $w_m$ are $m-$tensors in the $xy$ plane. Using the above 
Poisson summation formula, it is concluded that
$$
w_m(k,k_{y};r)=\sum_{n}D_m\psi_{n}(k,k_{y};r)-\frac{1}{2}
D_m H_{0}^{(1)}(k|r|)\,.
$$
The method to evaluate these series will be illustrated with the case $m=2$, while all technical 
details for the other needed cases will be omitted and only the final results will be stated.

As $|n|\to\infty$, 
$$
D_2 \psi_{n}(k,k_{y};r) \sim \sum_{-1\leq p\leq1}
\Big[\alpha^{+}_{p}(k,k_{y};x)\frac{1}{|n|^{p}}+\alpha^{-}_{p}(k,k_{y};x)\frac{s_n}{|n|^{p}}\Big]\phi_{n}(k,k_{y};r)
$$
where $s_n$ is the sign of $n$ and function $\phi_n$ are given by
\begin{eqnarray*}
\phi_{n}(k,k_{y};r)&=&e^{-k_{y,n}(|x|s_n -iy)}\,.
\end{eqnarray*}

The symmetric 2-tensor $\alpha_{p}^{\pm}$ is completely determined from the $n \rightarrow \infty$ limit of $D_{2}\psi_{n}$ for $|r| >0$. In order to calculate $w_{2}$  we only need the diagonal components of the tensor since the off-diagonal components are zero by symmetry:
\begin{align*}
(\alpha_{-1}^{+})_{xx} &=- (\alpha_{-1}^{+})_{yy}=  -2 \pi i, \\
(\alpha_{0}^{+})_{xx} &=-(\alpha_{0}^{+})_{yy}= -\frac{i k^{2} |x|}{2}, \\
(\alpha_{1}^{+})_{xx} &= (\alpha_{1}^{+})_{yy}= \frac{i k^{2}}{4 \pi},  \\
(\alpha_{0}^{-})_{xx} &= -(\alpha_{0}^{-})_{yy}=-i k_{y},
\end{align*}
and all other components are found to be zero.
Using this asymptotic behavior, the divergent part of the series in the limit $|r|\to 0^+$ 
can be identified. To this end, put $v=|x|+iy$ and
define the $2-$tensor  
\begin{eqnarray*}
P_2(k,k_{y};r)&=&\sum_{-1 \leq p \leq 1}[\alpha^{+}_{p}(k,k_{y};x)z_{p}^{+}(k_{y};r)+\alpha^{-}_{p}(k,k_{y};x)z_{p}^{-}(k_{y};r)]\,,\\
z_{p}^{+}(k_{y};r)&=&\sum_{n\neq 0}\frac{\phi_{n}(k,k_{y};r)}{|n|^{p}}=e^{-k_{y}v^*}Li_{p}(e^{-2\pi v^*})+
e^{k_{y}v}Li_{p}(e^{-2\pi v})\,,\\
z_{p}^{-}(k_{y};r)&=&\sum_{n\neq 0}\frac{\phi_{n}(k,k_{y};r)s_n}{|n|^{p}}=e^{-k_{y}v^* }Li_{p}(e^{-2\pi v^*})-
e^{k_{y} v}Li_{p}(e^{-2\pi v})
\end{eqnarray*}
where $Li_{p}(u)$ is the polylogarithm of order $p$. Then
\begin{eqnarray*}	
w_2(k,k_{y};r)&=& w_2^{\rm reg}(k,k_y;r)+ w_2^{\rm sin}(k,k_y;r)\,,\\
w_2^{\rm sin}(k,k_y;r)&=&P_2(k,k_{y};r)-\frac{1}{2}D_2H_{0}^{(1)}(k|r|)\,,\\
w_2^{\rm reg}(k,k_y;r)&=&\sum_{n}D_2\psi_{n}(k,k_{y};r)-P_2(k,k_{y};r)\\
&=&\sum_{n\neq 0}\Big\{D_2\psi_{n}-\Big(\alpha^{+}_{-1}|n|+\frac{|n|\alpha^{+}_{0}+n\alpha^{-}_{0}}{|n|}+
\frac{\alpha^{+}_{1}}{|n|}\Big)\phi_{n}\Big\}+D_{2}\psi_{0}\,,
\end{eqnarray*}

where the arguments $(k,k_y; r)$ in all functions were omitted for brevity. By definition of $P_2$,
divergent terms in the series for $w_2^{\rm reg}$ are cancelled for large $|n|$ and 
the series converges even for $r=0$. 
The distributional derivative $\sum_n \nabla_j\nabla_m\psi_n$ (and, hence, the distribution-valued tensor $P_2$)
is the sum of a singular part, that is equal to
 $-i \delta_{jm}\delta(r)$, and a regular distribution (being the corresponding classical derivative
wherever it exists).
The singular part exactly cancels with the singular part of the second distributional derivative of the Hankel function
in the expression for $w_2^{\rm sin}(k,k_y; r)$.
This cancellation occurs for all even-order derivatives. Therefore the limit $|r|\to 0^+$ can be computed
by studying the asymptotic behavior of the polylogarithm near its singular point. Recall that 
$Li_{p}(u)$ diverges as $u \rightarrow 1$ for $p\leq1$.
So, using the asymptotic form of $Li_p$ near its singular point and polar coordinates 
in the $xy$ plane,  one infers that
\begin{eqnarray*}
z_{1}^{+}(k_{y},r)  &=& -2 \ln{(2 \pi |r|)}+O(|r|)\,,
\\
z_{1}^{-}(k_{y},r) &=& \ln{(\frac{ |\cos{\theta}| +i \sin{\theta}}{|\cos{\theta}| -i \sin{\theta}})}+O(|r|)\,,
\\
z_{-1}^{+}(k_{y},r) &=& \frac{\cos{2 \theta}}{2 \pi^{2} |r|^{2}}-\frac{i k_{y} \sin{\theta}}{2 \pi^{2} |r|}+
\Big(\frac{k_{y}^{2}}{4 \pi^{2}}-\frac{1}{6}\Big)+O(|r|)\,,
\\
z_{-1}^{-}(k_{y},r) &=& \frac{i |\cos{\theta}| \sin{\theta}}{\pi^{2} |r|^{2}}-\frac{k_{y} |\cos{\theta}|}{2 \pi^{2} |r|} +O(|r|),
\\
z_{0}^{+}(k_{y},r) &=& \frac{|\cos{\theta}|}{\pi |r|}-1+O(|r|),
\\
z_{0}^{-}(k_{y},r) &=& \frac{i \sin{\theta}}{\pi |r|}-\frac{k_{y}}{\pi}+O(|r|),
\\
z_{2}^{+}(k_{y},r) &=& \frac{\pi^{2}}{3}+O(|r| \ln{|r|}),
\\
z_{-2}^{+}(k_{y},r) &=& \frac{|\cos{\theta}|(\cos^{2}{\theta}-3 \sin^{2}{\theta})}{2 \pi^{3} |r|^{3}}-\frac{i k_{y} 
|\cos{\theta}| \sin{\theta}}{\pi^{3} |r|^{2}}+\frac{k_{y}^{2}|\cos{\theta}|}{4 \pi^{3} |r|}+O(|r|)\,,\\
z_{-2}^{-}(k_{y},r)  &=& \frac{i \sin{\theta}(3 \cos^{2}{\theta} - \sin^{2}{\theta})}{2 \pi^{3} |r|^{3}}-\frac{k_{y} \cos{2 \theta}}{2 \pi^{3} |r|^{2}}+\frac{i k_{y}^{2} \sin{\theta}}{4 \pi^{3} |r|}-\frac{k_{y}^{3}}{12 \pi^{3}} +O(|r|),
\\
z_{3}^{+}(k_{y},r)  &=& 2 \zeta(3)+O(|r|^{2}\ln{|r|})\,, 
\end{eqnarray*}
$\theta$ is the polar angle counted from the positive $x$ axis counterclockwise and $\zeta(s)$ is 
the  Riemann zeta function. Using the above asymptotic equations and the asymptotic form 
of the Hankel function for a small argument, the limit of components $w^{\rm sin}_{2, jm}$
of the tensor $w^{\rm sin}_{2}$ is found

\begin{eqnarray*}
\lim_{r \rightarrow 0^{+}}w_{2,xx}^{\rm sin}(k,k_{y};r)
&=& \frac{ik^{2}}{2\pi}\ln{\frac{k}{4\pi}}+\frac{k^{2}}{4}
\Big[1+\frac{i(2\gamma-1)}{\pi}\Big]+\frac{ik_{y}^{2}}{2\pi}+\frac{i\pi}{3}\,,\\
\lim_{r \rightarrow 0^{+}}w_{2,yy}^{\rm sin}(k,k_{y};r)
&=& \frac{ik^{2}}{2\pi}\ln{\frac{k}{4\pi}}+\frac{k^{2}}{4}
\Big[1+\frac{i(2\gamma+1)}{\pi}\Big]-\frac{ik_{y}^{2}}{2\pi}-\frac{i\pi}{3}\,,\\
\lim_{r \rightarrow 0^{+}}w_{2,xy}^{\rm sin}(k,k_{y};r)&=&0\,,
\end{eqnarray*}

where $\gamma$ is the Euler constant. Therefore non-zero components of the tensor $W_2$ can be computed 
via the absolutely convergent series:

\begin{eqnarray*}
W_{2,xx}(k,k_{y})&=&-\sum_{n\neq 0}\Big(k_{x,n}-2\pi i|n|+\frac{ik^{2}}{4\pi|n|}-ik_{y}s_n\Big)\\
&&
-k_x+\frac{ik^{2}}{2\pi}\ln{\frac{k}{4\pi}}+\frac{k^{2}}{4}\Big[1+\frac{i(2\gamma-1)}{\pi}
\Big]+\frac{ik_{y}^{2}}{2\pi}+\frac{i\pi}{3}\,,\\
W_{2,yy}(k,k_{y})&=&-\sum_{n\neq 0}\Big(\frac{k_{y,n}^{2}}{k_{x,n}}+2\pi i |n| +\frac{ik^{2}}{4\pi |n|}+ik_{y}s_{n}\Big)
\\
&&-\frac{k_{y}^{2}}{k_x}
+ \frac{ik^{2}}{2\pi}\ln{\frac{k}{4\pi}}+\frac{k^{2}}{4}\Big[1+\frac{i(2\gamma+1)}{\pi}\Big]
-\frac{ik_{y}^{2}}{2\pi}-\frac{i\pi}{3}\,, 
\end{eqnarray*}

where $k_x=k_{x,0}$. 

%\textcolor{green}{Check the sign at the term with $\ln k$ by comparing it with 
%the limits for $w_2^{\rm sin}$}.

The same procedure can be used to evaluate the tensors $W_{{m}}(k,k_{y})$ for $m\neq 2$.  
One must first examine the asymptotic behavior of  the tensor $D_m\psi_n$ as $|n|\to \infty$.
This is accomplished by calculating 
the expansion 
$$
k_{x,n}^{p-1} (ik_{y,n})^q = \kappa_n^{(p,q)}+O\Big(n^{-3/2}\Big)
$$
where $p+q=m$.
The asymptotic coefficients 
$\kappa_n^{p,q}$ determine the tensor $P_m$ used to define $w_m^{\rm reg}$ and $w_m^{\rm sin}$.
The cancellation of the singular distributional part in the tensor $w_m^{\rm sin}$ is then established,
and the asymptotic properties of the polylogarithm function near its singular point are exploited
to find the limit of $w_m^{\rm sin}$ as $|r|\to 0^{+}$ as an absolutely convergent series
$$
W_m(k,k_y)=\sum_{n\neq 0}\Big(k_{x,n}^{p-1} (ik_{y,n})^q-\kappa_{n}^{(p,q)}\Big) +C_{m,pq}
$$
where $p$ and $q$ are the numbers of the $x$ and $y$ derivatives in $D_m$, respectively, and 
the series arises from the limit of $w_m^{\rm reg}$, while the constant $C_{m,pq}$ is the 
sum of $k_x^{p-1}(ik_y)^q$ and the limit
of $w_m^{\rm sin}$ obtained by the asymptotic expansion of the polylogarithm and Hankel functions.

It should be noted that all series involving an odd number of $x$ derivatives will be zero from the parity symmetry of the series. Here are the results
 up to $m=4$ that are needed to define the scattering amplitudes for the single and double array.
In particular, representations of the components of tensors $W_m$ via absolutely convergent series 
are essential when their calculating numerical values for given $k$ and $k_y$.  
For $m=0,1$ 

\begin{eqnarray*}
W(k,k_{y})&=&
\sum_{n\neq 0}\Big(\frac{1}{k_{x,n}}+\frac{i}{2\pi |n|}\Big)+\frac{1}{k_x}-
\frac{i}{\pi}\Big(\ln{\frac{k}{4\pi}}+\gamma\Big)-\frac{1}{2}\,,\\
W_{1,y}(k,k_{y})&=&\sum_{n\neq 0}\Big(\frac{ik_{y,n}}{k_{x,n}}-s_n\Big)+\frac{ik_{y}}{k_{x}}-\frac{k_{y}}{\pi}\,.
\end{eqnarray*}

%\textcolor{green}{Check the constant term in $W_{1,y}$.}
For $m=3$

\begin{eqnarray*}
W_{3,xxy}(k,k_{y})&=&\sum_{n\neq 0}\Big(-ik_{y,n}k_{x,n}-4\pi^{2} |n| n -4\pi k_{y}|n|
- (k_{y}^{2}-{\textstyle\frac{1}{2}}k^2)s_n
\Big)\\
&&
-ik_{y}k_x+\frac{k_{y}k^{2}}{2\pi}-\frac{k_{y}^{3}}{3 \pi}-\frac{2 \pi k_{y}}{3}\,,\\
W_{3, yyy}(k,k_{y})&=&\sum_{\mid m\mid>0}\Big(\frac{-ik_{y,m}^{3}}{k_{x,n}}+4 \pi^{2} |n| n
+4\pi k_{y}|n|
+(k_{y}^{2}+{\textstyle\frac{1}{2}}k^2)s_n\Big)\\
&&
-\frac{i k_{y}^{3}}{k_x}+\frac{k_{y}k^{2}}{2\pi}+\frac{k_{y}^{3}}{3 \pi}+\frac{2 \pi k_{y}}{3}\,.
\end{eqnarray*}

For $m=4$,

\begin{eqnarray*}
W_{4,xxxx}(k,k_{y})&=&\sum_{n\neq 0}\Big(k_{x,n}^3-\kappa_n^{(4,0)}\Big)+C_{4,xxxx}\,,\\
\kappa_n^{(4,0)}&=&-i\Big[(2\pi|n|)^3+
12 \pi^{2}k_{y} |n| n+\pi(6 k_{y}^{2}-3k^{2})|n|
+(k_{y}^{3}-{\textstyle\frac{3}{2}} k^{2}k_{y})s_n+\frac{3k^{4}}{16\pi |n|}\Big]\\
C_{4,xxxx}&=&k_x^3-\frac{3 k^{4}}{16}+
i\Big[\pi k_{y}^{2}-\frac{\pi k^{2}}{2}-\frac{3 k^{2} k_{y}^{2}}{4 \pi}-
\frac{2 \pi^{3}}{15}+\frac{k_{y}^{4}}{4 \pi}-\frac{3 k^{4}}{8 \pi}
\Big(\ln{\frac{k}{4 \pi}}+\gamma-\frac{3}{4}\Big)\Big]\,,\\
W_{4,yyyy}(k,k_{y})&=&\sum_{n\neq 0}\Big(\frac{k_{y,n}^{4}}{k_{x,n}}-\kappa_n^{(0,4)}\Big)+C_{4,yyyy}\,,\\
\kappa_n^{(0,4)}&=&-i\Big[(2\pi |n|)^{3}+12\pi^{2}k_{y}|n| n+\pi(k^{2}+6k_{y}^{2})|n|
+(k_{y}^{3}+{\textstyle\frac{1}{2}}k^{2}k_{y})s_n+\frac{3k^{4}}{16\pi |n|}\Big]\\
C_{4,yyyy}&=&\frac{k_{y}^{4}}{k_x}-\frac{3 k^{4}}{16}+i\Big[\frac{k^{2}k_{y}^{2}}{4 \pi}+\frac{\pi k^{2}}{6}+\frac{k_{y}^{4}}{4 \pi}+\pi k_{y}^{2}-\frac{2 \pi^{3}}{15}-\frac{k^{4}}{8}
\Big(3\ln{\frac{k}{4\pi}}+3\gamma+\frac{7}{4}\Big)\Big]\,,\\
W_{4,xxyy}(k,k_{y})&=&\sum_{n\neq 0}\Big(k_{y,n}^{2}k_{x,n}-\kappa_{n}^{(2,2)}\Big)+C_{4,xxyy}\,,\\
\kappa_{n}^{(2,2)}&=&i\Big[(2\pi |n|)^{3}+12\pi^{2}k_{y}|n| n+\pi(6k_{y}^{2}-k^{2})|n| 
+(k_{y}^{3}-{\textstyle\frac{1}{2}}k^{2}k_{y})s_n-\frac{k^{4}}{16\pi |n|}\Big]\,,\\
C_{4,xxyy}&=&k_{y}^{2}k_x-\frac{k^{4}}{16}+
i\Big[\frac{k^{2}k_{y}^{2}}{4\pi}+\frac{\pi k^{2}}{6}-
\frac{k_{y}^{4}}{4 \pi}-\pi k_{y}^{2}+\frac{2 \pi^{3}}{15}-\frac{k^{4}}{8 \pi}
\Big(\ln{\frac{k}{4 \pi}}+\gamma+\frac{1}{4}\Big)\Big].
\end{eqnarray*}

% \textcolor{green}{1. Check the sign of $W_{4,xxyy}$ or the sign of the series in it.\\
%2. Express the tensors $\Pi(0)$, $D\Pi(0)$, and $D^2\Pi(0)$ in terms of $W_m$ if it is 
%not too complicated, otherwise give $\Pi(0)$ in terms of $W$ and $W_2$ as an example}

In order to evaluate the tensors $\Pi(0)$, $D\Pi(0)$, and $D^2\Pi(0)$ we also require the following integrals
\begin{equation*}
\int_{|r| < R}G_{ij}(r)d^{2}r
\end{equation*}
as well as
\begin{equation*}
\int_{|r| < R}G_{ij,mn,l}(r)d^{2}r,
\end{equation*}
and
\begin{equation*}
\int_{|r| < R}G_{ij,mn}(r)d^{2}r.
\end{equation*}
Using the following identity for cylindrical harmonics $(C_{n}(k;r) = H_{n}^{(1)}(k|r|)e^{i n \theta})$
\begin{align*}
\nabla_{x}C_{n}(k;r) = \frac{k}{2}(C_{n-1}(k;r)-C_{n+1}(k;r)), \\
\nabla_{y}C_{n}(k;r) = \frac{ik}{2}(C_{n-1}(k;r)+C_{n+1}(k;r)),
\end{align*}
as well as the integral 
\begin{equation*}
\int_{|r|<R}C_{n}(k;r)d^{2}r = \delta_{n0} \int_{|r|<R}C_{0}(k;r)d^{2}r = \kappa_{0}(k,R) \delta_{n0},
\end{equation*}
where
\begin{eqnarray*}
\kappa_{0}(k,R) &=& \int_{|r|<R}C_{0}(k;r)d^{2}r = \frac{2 \pi \delta_{n0}}{k}\lim_{\Delta R \rightarrow 0^{+}} r H_{1}^{(1)}(kr)|_{r = \Delta R}^{R}\\
&=& \pi R^{2}[1+\frac{i}{\pi}(2 \ln{(\frac{k R}{2})} + 2\gamma - 1)] + O(R^{4}\ln{R}).
\end{eqnarray*}
Given the above formula one can trivially evaluate the relevant integrals
\begin{align*}
\int_{|r| < R}\nabla_{i}^{2}H_{0}^{(1)}(k|r|)d^{2}r =-\frac{k^{2}}{2}\kappa_{0}(k,R), \\
\int_{|r| < R}\nabla_{i}^{4}H_{0}^{(1)}(k|r|)d^{2}r = \frac{3k^{4}}{8}\kappa_{0}(k,R), \\
\int_{|r| < R}\nabla_{x}^{2}\nabla_{y}^{2}H_{0}^{(1)}(k|r|)d^{2}r = \frac{k^{4}}{8}\kappa_{0}(k,R),
\end{align*}
for $i \in \{x,y\}$. It's clear that all terms with an odd number of partial derivative in both variables integrates out to $0$ due to the parity symmetry. Now one can easily evaluate $\Pi(0)$, $D\Pi(0)$, and $D^2\Pi(0)$:
\begin{align*}
\Pi_{xx}(0) &= \frac{i}{2 \omega^{2}}[\epsilon(W_{2,xx}(k_{l},k_{y})+W_{2,yy}(k_{t},k_{y}))-\frac{k_{l}^{2}}{4}\kappa_{0}(k_{l},R)-\frac{k_{t}^{2}}{4}\kappa_{0}(k_{t},R)],\\
\Pi_{yy}(0) &= \frac{i}{2 \omega^{2}}[\epsilon(W_{2,xx}(k_{t},k_{y})+W_{2,yy}(k_{l},k_{y}))-\frac{k_{l}^{2}}{4}\kappa_{0}(k_{l},R)-\frac{k_{t}^{2}}{4}\kappa_{0}(k_{t},R)],
\end{align*}
\begin{align*}
\Pi_{xx,y}(0) &= \frac{i\epsilon}{2 \omega^{2}}(W_{3,xxy}(k_{l},k_{y})+W_{2,yyy}(k_{t},k_{y})), \\
\Pi_{yy,y}(0) &= \frac{i\epsilon}{2 \omega^{2}}(W_{3,xxy}(k_{t},k_{y})+W_{2,yyy}(k_{l},k_{y})), \\
\Pi_{xy,x}(0) &= \frac{i\epsilon}{2 \omega^{2}}(W_{3,xxy}(k_{l},k_{y})-W_{2,yyy}(k_{t},k_{y})), 
\end{align*}
\begin{align*}
\Pi_{xx,yy}(0) &= \frac{i}{2 \omega^{2}}[\epsilon(W_{4,xxyy}(k_{l},k_{y})+W_{4,yyyy}(k_{t},k_{y}))+\frac{k_{l}^{4}}{16}\kappa_{0}(k_{l},R)+\frac{3 k_{t}^{4}}{16}\kappa_{0}(k_{t},R)],\\
\Pi_{yy,yy}(0) &= \frac{i}{2 \omega^{2}}[\epsilon(W_{4,xxyy}(k_{t},k_{y})+W_{4,yyyy}(k_{l},k_{y}))+\frac{k_{t}^{4}}{16}\kappa_{0}(k_{t},R)+\frac{3 k_{l}^{4}}{16}\kappa_{0}(k_{l},R)],\\
\Pi_{xx,xx}(0) &= \frac{i}{2 \omega^{2}}[\epsilon(W_{4,xxxx}(k_{l},k_{y})+W_{4,xxyy}(k_{t},k_{y}))+\frac{3k_{l}^{4}}{16}\kappa_{0}(k_{l},R)+\frac{ k_{t}^{4}}{16}\kappa_{0}(k_{t},R)],\\
\Pi_{yy,xx}(0) &= \frac{i}{2 \omega^{2}}[\epsilon(W_{4,xxxx}(k_{t},k_{y})+W_{4,xxyy}(k_{l},k_{y}))+\frac{3k_{t}^{4}}{16}\kappa_{0}(k_{t},R)+\frac{ k_{l}^{4}}{16}\kappa_{0}(k_{l},R)],\\
\Pi_{xy,xy}(0) &= \frac{i}{2 \omega^{2}}[\epsilon(W_{4,xxyy}(k_{l},k_{y})-W_{4,xxyy}(k_{t},k_{y}))+\frac{ k_{l}^{4}}{16}\kappa_{0}(k_{l},R)-\frac{ k_{t}^{4}}{16}\kappa_{0}(k_{t},R)].
\end{align*}

The above equations along with the interface conditions (continuity of field and normal traction) completely determine both the spectrum as well as the transmission and reflection coefficients, hence the scattering problem is completely solved in principle.

\end{document}